\documentclass[12pt]{article}
\usepackage{amsmath} 
\usepackage{multirow} 
\usepackage{amsfonts}
\usepackage{amssymb,graphics,psfrag}
\usepackage{array,epsfig,multirow,stmaryrd,graphicx}
\usepackage{comment}
\def\hybrid{\topmargin -20pt    \oddsidemargin 0pt
        \headheight 0pt \headsep 0pt
        \textwidth 6.25in       
        \textheight 9 in       
        \marginparwidth .875in
        \parskip 5pt plus 1pt 
          \jot = 1.5ex
   }
\hybrid
\numberwithin{equation}{section}
\numberwithin{table}{section}

\newcommand{\beq}{\begin{equation}}
\newcommand{\eeq}{\end{equation}}
\newcommand{\be}{\begin{equation}}
\newcommand{\ee}{\end{equation}}
\newcommand{\bea}{\begin{eqnarray}}
\newcommand{\eea}{\end{eqnarray}}   
\newcommand{\ben}{\begin{eqnarray*}}
\newcommand{\een}{\end{eqnarray*}}                  
\newcommand{\ba}{\begin{aligned}}
\newcommand{\ea}{\end{aligned}}
\newcommand{\bt}{\begin{tabular}}
\newcommand{\et}{\end{tabular}}
\newcommand{\bc}{\begin{center}}
\newcommand{\ec}{\end{center}}

\newcommand{\cT}{\mathcal{T}}

\newcommand{\cK}{\mathcal{K}}
\newcommand{\cN}{\mathcal{N}}
\newcommand{\cW}{\mathcal{W}}

\newcommand{\cF}{\mathcal{F}}

\newcommand{\cM}{\mathcal M}

\newcommand{\I}{\text{Im}}
\newcommand{\R}{\text{Re}}

\newcommand{\bbZ}{\mathbb{Z}}
\newcommand{\bbR}{\mathbb{R}}

\newcommand{\bbP}{\mathbb{P}}

\newcommand{\nn}{\nonumber}

\newcommand{\cref}{{\bf [check ref]}}

\newcommand{\simga}{\sigma}

\def\IR{{\mathbb R}}

\psfrag{n1}{$\nu_1$}
\psfrag{n2}{$\nu_2$}
\psfrag{n1'}{$\nu_1'$}
\psfrag{n2'}{$\nu_2'$}
\psfrag{n9}{$\nu_9$}
\psfrag{n10'}{$\nu_{10}'$}
\psfrag{t1}{$\tilde{\nu}_1$}
\psfrag{t2}{$\tilde{\nu}_2$}
\psfrag{t9}{$\tilde{\nu}_9$}
\psfrag{t1'}{$\tilde{\nu}_1'$}
\psfrag{t2'}{$\tilde{\nu}_2'$}
\psfrag{t10'}{$\tilde{\nu}_{10}'$}

\newdimen\tableauside\tableauside=1.0ex
\newdimen\tableaurule\tableaurule=0.4pt
\newdimen\tableaustep
\def\phantomhrule#1{\hbox{\vbox to0pt{\hrule height\tableaurule width#1\vss}}}
\def\phantomvrule#1{\vbox{\hbox to0pt{\vrule width\tableaurule height#1\hss}}}
\def\sqr{\vbox{%
  \phantomhrule\tableaustep
  \hbox{\phantomvrule\tableaustep\kern\tableaustep\phantomvrule\tableaustep}%
  \hbox{\vbox{\phantomhrule\tableauside}\kern-\tableaurule}}}
\def\squares#1{\hbox{\count0=#1\noindent\loop\sqr
  \advance\count0 by-1 \ifnum\count0>0\repeat}}
\def\tableau#1{\vcenter{\offinterlineskip
  \tableaustep=\tableauside\advance\tableaustep by-\tableaurule
  \kern\normallineskip\hbox
    {\kern\normallineskip\vbox
      {\gettableau#1 0 }%
     \kern\normallineskip\kern\tableaurule}%
  \kern\normallineskip\kern\tableaurule}}
\def\gettableau#1{\ifnum#1=0\let\next=\null\else
\squares{#1}\let\next=\gettableau\fi\next}

\def\blfootnote{\xdef\@thefnmark{}\@footnotetext} 
\long\def\symbolfootnote[#1]#2{\begingroup%
\def\thefootnote{\fnsymbol{footnote}}\footnote[#1]{#2}\endgroup}

\begin{document}

\baselineskip=16pt

\begin{titlepage}
\begin{flushright}
\parbox[t]{1.3in}{
MAD-TH-07-07\\
0705.3253\ [hep-th]}
\end{flushright}

\begin{center}

\vspace*{ 1.5cm}

{\large \bf  Non-Perturbative Corrections and Modularity \\[.4cm]
 in $\cN=1$ Type IIB Compactifications}

\vskip 2cm

\begin{center}
 \bf{Thomas W.~Grimm} \footnote{grimm@physics.wisc.edu        
}
\end{center}

{\em Department of Physics, University of Wisconsin, \\[.1cm]
        Madison, WI 53706, USA}
\vspace{.3cm}

 \vspace*{1cm}

\end{center}

\vskip 0.5cm

\begin{center} {\bf ABSTRACT } \end{center}

\vskip .4cm

Non-perturbative corrections  and modular properties of four-dimensional 
type IIB Calabi-Yau orientifolds are discussed. It is shown that certain 
non-perturbative $\alpha'$ corrections survive in the 
large volume limit of the orientifold and periodically correct the K\"ahler potential. 
These corrections depend on the NS-NS two form and have to be completed 
by D-instanton contributions 
to transform covariantely under 
symmetries of the type IIB orientifold background.
It is shown that generically 
also the D-instanton superpotential depends on the two-form moduli as well as
on the complex dilaton. These contributions can arise through
theta-functions
with the dilaton as modular parameter. An orientifold of the Enriques Calabi-Yau
allows to illustrate these general considerations.
It is shown that this compactification leads to  a controlled four-dimensional $\cN=1$ effective theory
due to the absence of various quantum corrections.
Making contact to the underlying topological 
string theory the D-instanton superpotential is proposed to be related to a specific
modular form counting D3, D1, D(-1) degeneracies on the Enriques 
Calabi-Yau.

\vskip .4cm
\hfill May, 2007
\end{titlepage}

\section{Introduction}

Recently much effort has focused on the study 
of orientifold compactifications of type II string theory
with space-time filling D-branes and background fluxes. 
The reason is that these compactifications can lead to calculable 
four-dimensional effective theories supporting string vacua relevant for 
particle physics and cosmology \cite{reviewPP,reviewcosmo,FluxReviews}.
Particularly well controlled  are warped type IIB Calabi-Yau orientifolds with
space-time filling D3 and D7 branes which yield a four-dimensional effective
theory with $\cN=1$ supersymmetry \cite{GKP,FluxReviews}.
It was realized that in these compactifications 
the inclusion of background fluxes and certain non-perturbative  
corrections might lead to a stabilization of all unwanted scalar moduli 
fields in a local vacuum \cite{KKLT}. This was 
demonstrated for specific examples e.g.~in refs.~\cite{DDFGK,BBCQ,Lust1,Lust2} and strengthened the 
believe in a vast landscape of supersymmetric and non-supersymmetric 
string vacua \cite{FluxReviews}. In order to study these vacua 
a precise knowledge of the $\cN=1$ characteristic data of the four-dimensional 
effective theory is of central importance. In particular, this includes the
understanding of perturbative and non-perturbative corrections 
to the K\"ahler potential and the superpotential.

The aim of this work is to investigate the leading perturbative and 
non-perturbative corrections for Calabi-Yau orientifolds with 
O3 and O7 planes. 
We first study the 
$\alpha'$ corrections inherited from the underlying $\cN=2$
theory which survive the large volume limit of the orientifold. 
This includes the perturbative $\alpha'$ corrections discussed
in ref.~\cite{BBHL}. Moreover, we  argue by using the results of refs.~\cite{GL1,GL2} 
that also non-perturbative $\alpha'$ corrections involving the 
NS-NS B-field can survive the large volume limit of the orientifold. 
These corrections are generically present in compactifications in which the 
B-field is not entirely projected out by the orientifold symmetry.%
\footnote{An example of a Calabi-Yau orientifold with non-vanishing B-field 
moduli is presented in the second part of this paper. For other 
examples which admit these additional moduli fields, see e.g.~ref.~\cite{Lust2}.}
The real B-field scalars combine with the scalars of the R-R 
two-form $C_2$ into complex scalars $G^a$ through the combination $C_2-\tau B_2$,
where $\tau$ is the complex dilaton-axion \cite{GL1}.
The perturbative and non-perturbative $\alpha'$ corrections in the orientifold
large volume limit do not correct the $\cN=1$ coordinates.  
They do however contribute to the K\"ahler potential and 
we will be able to determine these corrections explicitly 
in terms of the topological invariants of the underlying Calabi-Yau manifold.
We will also study the non-perturbative superpotential generated by D3-instantons
wrapping a four-cycle in the Calabi-Yau manifold and show that it generically 
depends on the scalars $\tau$ and $G^a$. In order to 
do that, we implement the non-perturbative symmetries
inherited from the full type IIB string theory.

Type IIB string theory possesses a strong-weak duality known as S-duality. 
This non-perturbative symmetry relates one type IIB theory with complex 
string coupling $\tau$ to a dual type IIB string theory with string coupling $-1/\tau$.
Moreover, it exchanges the NS-NS and R-R two-forms and thus fundamental
strings with D1 branes. Together with the shifts in the axion, $\tau \rightarrow \tau+1$, the S-duality transformation 
generates the discrete duality group $Sl(2,\bbZ)$. In an $\cN=1$ compactification
this group will generically be reduced further or broken completely by
to the non-trivial background geometry. 
However, in the orientifold compactifications under consideration the complex dilaton $\tau$
does not vary over the compact six-dimensional geometry and appears as four-dimensional chiral 
field \cite{GKP, FluxReviews}. In this limit we expect that a subgroup $\Gamma_S$
of the full $Sl(2,\bbZ)$ duality is a symmetry of the four-dimensional theory in 
analogy to refs.~\cite{FILQ,CFILQ}. Determining the transformations of the $\cN=1$ coordinates under 
 $\Gamma_S$ as well as integral shifts of the NS-NS B-field allows us to 
 study the moduli dependence and symmetries of the K\"ahler potential and 
 superpotential in the orientifold large volume limit. 
 
 We begin by discussing the transformation properties of the K\"ahler potentials under
$\Gamma_S$ when $\alpha'$ corrections are included. In order for these to be invariant under 
$\Gamma_S$ also contributions from D1 and D(-1) branes
have to be taken into account. In general, it is hard to compute these 
corrections. We will however be able to discuss candidate 
completions which reproduce the perturbative and non-perturbative 
$\alpha'$ corrections and admit the desired transformation properties.
In order to obtain these solutions we will simply sum over images 
of the $\alpha'$ corrections under the duality group following \cite{GG, RRSTV}. 
This does however not guarantee that the result is the true non-perturbative 
completion. Firstly, this analysis is only valid in the orientifold limit in which 
the type IIB symmetry is not entirely broken by the vacuum and a discrete 
group $\Gamma_S$ is preserved. Secondly, even though this symmetry group 
ideally restricts the answer to be generated by a finite set of appropriately 
transforming functions additional boundary conditions are needed to
fix the precise form of the duality invariant completion.\footnote{See \cite{BCOV,YY,Klemm, GKMW} for the discussion of 
an analogous problem within topological 
string theory.}
For corrections to the $\cN=1$ K\"ahler potential this task
is even more involved, since the K\"ahler potential is not
protected by holomorphicity or non-renormalization theorems.
The application of string-string dualities such as heterotic-F-theory duality 
might however help to compute these corrections explicitly as 
argued, for example, in refs.~\cite{BM, HMS}.
One expects that modularity arguments are however more powerful 
when arguing about the superpotential. 

In $\cN=1$ theories the superpotential is  holomorphic  and protected 
against perturbative corrections. For the type IIB orientifold 
setups the determination of the D3-instanton superpotential
is of central importance. However, its explicit form is 
in general hard to determine \cite{Witten,DGW, CL}.
Nevertheless, by
combining holomorphicity and modular properties under the inherited 
type IIB $Sl(2,\bbZ)$ symmetry as well as shifts in the 
NS-NS B-field the moduli dependence of the superpotential
in general large volume orientifolds can be discussed. In case 
the complex dilaton $\tau$ varies over the internal space only a local 
analysis of the superpotential can be performed \cite{Witten:1996hc,Ganor}. Here 
our results are more restrictive due to the fact that $\tau,G^a$ do 
not vary over the compact space. 
We find that the
complex fields $G^a$ depending on the NS-NS and R-R two-form
moduli naturally arise through products of theta-functions
and modular forms with the complex 
dilaton-axion $\tau$ as modular parameter. In the second part of the paper 
we propose that this set of theta-functions can be determined for 
a specific orientifold example.

The specific example we consider is an orientifold of the Enriques 
Calabi-Yau. The underlying Calabi-Yau manifold is a $K3$ fibration of the form 
$Y_E=(K3 \times T^2)/\bbZ_2$ \cite{Borcea,FHSV}, where the freely acting $\bbZ_2$ symmetry 
yield a minus sign on the complex coordinate of $T^2$ and acts as the Enriques involution on the $K3$ surface \cite{SurfaceB}.
We will show that an appropriate definition of the 
orientifold projection allows to explicitly determine the $\cN=1$ four-dimensional 
effective theory. Since the geometric moduli space of the 
underlying $\cN=2$ theory is not corrected by world-sheet instantons or perturbative 
$\alpha'$ corrections the resulting $\cN=1$ theory is particularly well controlled.
We will show that the $\cN=1$ moduli space is a product of two cosets $\tilde \cM_{\rm sk} \times \tilde \cM_{\rm q}$.
The first factor $\tilde \cM_{\rm ks}$ arises from the reduction of the
$\cN=2$ special K\"ahler manifold containing the complex structure deformations of $Y_E$. It is itself a 
special K\"ahler manifold and was studied intensively in the literature \cite{KM,GKMW}.
The reduction of the $\cN=2$ quaternionic manifold leads to a K\"ahler manifold 
$\tilde \cM_{\rm q}$ of half its dimension. Remarkably, 
$\tilde \cM_{\rm q}$ can be identified with the original $\cN=2$ special 
K\"ahler manifold of complexified K\"ahler structure deformations $\cM_{\rm ks}$ times 
an $Sl(2,\bbR)/U(1)$ factor. 
In this identification half of the 
NS-NS fields arising as real parts of coordinates on $\cM_{\rm sk}$ are
replaced by R-R fields. The resulting $\cN=1$ coordinates encode the 
correct couplings  to D(-1), D1 and D3 branes. Note however, that 
this duality is not performed in the large volume coordinates on $\cM_{\rm sk}$,
but rather at a special locus where also the volume of the K3 fiber can be 
small. 

The physics in the regime where the K3 fiber of the Enriques Calabi-Yau is 
small was studied intensively in the underlying $\cN=2$ theory. It was shown 
in ref.~\cite{aspinwall} that at the limit were the K3 fiber is of Planck length the type II 
theory undergoes a phase transition somewhat similar to the well-known 
conifold transition. It was later argued in ref.~\cite{KM} that the light BPS
degrees of freedom at this locus are bound states of D4, D2 and D0
branes wrapped around specific four and two-cycles of $Y_E$. 
The authors of \cite{KM} showed that the topological string theory on the Enriques
Calabi-Yau can be resummed to count the degeneracies of these
degrees of freedom. The leading contributions arise through a particular
holomorphic function $\Phi_{\rm B}$ known from the work of Borcherds \cite{borcherdsone, borcherds} 
and Harvey, Moore \cite{HM,HM2}. Here we will employ the duality of the theory on
$\cM_{\rm ks}$ at this special locus to the corresponding orientifold theory.
We propose that $\Phi_{\rm B}$ naturally arises in the $\cN=1$ superpotential containing
the D3-instanton corrections proportional to $e^{i T_S}$, where $T_S$ contains the 
volume of the K3 fiber. In accord with our general considerations, 
the coefficients are indeed generalizations of theta functions 
depending on the modular parameter $\tau$, the dilaton-axion, as well as the scalars 
$G^a$ arising from the NS-NS 
and R-R two-forms. The study of the Enriques orientifold exemplifies nicely 
the interplay of holomorphicity and symmetry properties for the non-perturbative 
superpotential. 

This paper is organized as follows. In section \ref{orirev} we briefly review the effective
theory of type IIB orientifolds with O3 and O7  planes. We discuss the reduction 
of an $\cN=2$ theory defined by two general pre-potentials for complex structure 
and K\"ahler structure deformations respectively. It is then shown in section \ref{largevolume}
that certain $\alpha'$ corrections survive in the large volume limit of the orientifold 
and correct the K\"ahler potential in an explicitly calculable way.  
The modular completion of these corrections by D(-1) and D1 brane contributions 
is discussed in section \ref{Kaehlersymm}. In section \ref{generalsup} we turn to the discussion 
of the non-perturbative superpotential generated by D3-instantons. We study its 
transformations under the type IIB symmetries and argue for a moduli dependence 
through generalizations of theta functions. In section \ref{EnriquesO} we present an explicit example
by introducing an orientifold of the Enriques Calabi-Yau manifold. We first summarize 
some details about the $\cN=2$ theory in section \ref{GeometryN=2}. The K\"ahler potential and 
an interesting duality map is studied in \ref{effectiveN=1action}. Finally, in section \ref{EnriquesW} we propose 
a particular non-perturbative superpotential counting degeneracies of D3, D1, D(-1)
bound states.

\section{Non-perturbative Corrections and Modularity}

In this section we discuss non-perturbative corrections and 
the transformation properties of the $\cN=1$ effective action 
of type IIB string theory compactified on an orientifold background. We begin 
with a brief review of the four-dimensional effective theory in section 
\ref{orirev}. In section \ref{largevolume} we show that in the orientifold large volume 
limit the perturbative and certain non-perturbative $\alpha'$ corrections
inherited from the underlying $\cN=2$
theory correct the $\cN=1$ K\"ahler potential. 
We will argue that these corrections generically do not respect the 
type IIB $Sl(2,\bbZ)$ symmetries in section \ref{Kaehlersymm}. Since in the orientifold limit
a subgroup
$\Gamma_S$ of this symmetry group is expected to 
be preserved we comment on modular completions of the K\"ahler potential.
Finally, in section \ref{generalsup} we analyze the transformation properties of the $\cN=1$ complex 
coordinates and constrain the D-instanton superpotentials to contain generalizations 
of theta functions. This leads to a new moduli dependence of the superpotential 
which is generic for many orientifold compactifications.

\subsection{Brief review of the effective action of type IIB orientifolds
 \label{orirev}}

In this section we review the $\cN=1$ effective supergravity theory arising 
by compactification of type IIB supergravity on an orientifold background
following \cite{GKP,BBHL,GL1,GL2,TG}. We will focus on orientifold projections yielding O3 and O7 planes
and include the leading perturbative $\alpha'$ corrections \cite{BBHL} as well as the world-sheet instanton 
corrections inherited from the underlying $\cN=2$ theory \cite{GL2}. Since 
there exists a number of reviews \cite{FluxReviews} on this topic we will keep our discussion brief.

In type IIB orientifolds with O3/O7 planes the orientifold projection takes 
the form $(-1)^{F_L} \Omega_p \sigma$, where $F_L$ is the left fermion 
number, $\Omega_p$ is the world-sheet parity reversal and $\sigma$ is
some geometric involutive symmetry of the background. 
In order to preserve $\cN=1$ supersymmetry 
$\simga$ has to be a holomorphic and isometric involution. It acts
non-trivially on the internal Calabi-Yau manifold $Y$ and leaves the four flat directions invariant.
For models with O3/O7 planes $\sigma$ acts on the K\"ahler form $J$ and  holomorphic three form $\Omega$ of $Y$
as
\beq \label{JOmegatrans}
   \sigma^* J = J \ , \qquad \sigma^* \Omega = - \Omega\ ,
\eeq
where $\sigma^*$ is the pull-back.
In order to remain in the spectrum the NS-NS and R-R fields have to transform as
follows under $\sigma^*$. The dilaton $\phi$, the axion $C_0$ as well as the 
four-form $C_4$ are invariant under the action of $\sigma$, while the NS-NS two-form $B_2$ and R-R 
two-form $C_2$ transform with a minus sign.
Type IIB Calabi-Yau orientifolds with O3/O7 planes have the following truncated $\cN=1$ moduli 
space:
\beq \label{modspace1}
   \tilde \cM_{\rm sk} \times \tilde \cM_{\rm q}\ ,
\eeq
where $ \tilde \cM_{\rm sk}$ is a special K\"ahler manifold inside the $\cN=2$ special K\"ahler 
manifold $\cM_{\rm sk}$ and $\tilde \cM_{\rm q}$ is a K\"ahler manifold inside the $\cN=2$ quaternionic 
manifold $\cM_{\rm q}$. In the following we will describe the geometry of the moduli 
space \eqref{modspace1} in more detail.

Let us start with some comments on the cohomology of the orientifold 
theory and the reduction of $\cM_{\rm sk}$.  Since $\sigma$ is a holomorphic
involution the cohomology groups $H^{(p,q)}$ split
into two eigenspaces under the action of $\sigma^*$ as
$H^{(p,q)}=H^{(p,q)}_+ \oplus H^{(p,q)}_-$.   
We denote the dimensions of $H^{(p,q)}_\pm$ by $h^{(p,q)}_\pm$.
The four-dimensional invariant spectrum is found by using a Kaluza-Klein 
expansion in harmonic forms keeping only the fields which in addition 
obey the correct transformations under $\sigma^*$. This induces a reduction 
of the special K\"ahler manifold $\cM_{\rm sk}$
for the orientifold setups.
Since $\sigma$  transforms the complex three-form $\Omega$ 
with a minus sign the complex structure deformations parametrized 
by the elements of $H^{(2,1)}$ are reduced to $h^{(2,1)}_-$  complex 
scalars $z^k$. It can be shown that these define a $h^{(2,1)}_-$ dimensional 
special K\"ahler submanifold $\tilde \cM_{\rm sk}$ of the original $\cN=2$ moduli 
space of complex structure deformations. The K\"ahler potential on $\tilde \cM_{\rm sk}$
takes the well-known form 
\beq
 K_{\rm cs}(z,\bar z) =-\ln\big[ i \int_Y \Omega(z) \wedge \bar \Omega(\bar z)\big] \ ,
\eeq
where $\Omega(z^k)$ varies holomorphically  over $\tilde \cM_{\rm sk}$. 
Recall that in the underlying $\cN=2$ theory the complex scalars $z$ were 
part of vector multiplets. In the orientifold reduction also $h^{(2,1)}_+$ of the 
vectors survive. The gauge-kinetic coupling function is the second derivative
of the pre-potential of the underlying $\cN=2$ special K\"ahler manifold $\cM_{\rm sk}$
with respect to the complex structure deformations $z^\kappa$, which are then set to zero 
in the orientifold scenario \cite{GL1}.

The reduction of the quaternionic space $\cM_{\rm q}$ is slightly more involved. 
Since $\sigma$  leaves the K\"ahler
form $J$ invariant and yields a minus sign 
on the $B_2$ field we expand
\beq \label{JBexpansion}
   J = v^\alpha \omega_\alpha \ ,\quad \alpha=1,\ldots,h^{(1,1)}_+\ , \qquad \qquad B_2 = b^a \omega_a\ , \quad a=1,\ldots,h^{(1,1)}_-\ ,
\eeq
where $\omega_\alpha$ is an integral basis of $H^2_+(Y,\bbZ)$ and  $\omega_a$ is an integral basis of $H^2_-(Y,\bbZ)$.
The conditions \eqref{JBexpansion} defines a real subspace of the $h^{(1,1)}$ dimensional 
space of complexified K\"ahler deformations $\cM_{\rm ks}$ of $Y$. This is due to the fact that either the 
real or the complex part of the complexified K\"ahler form survives:
\beq \label{complexJ}
   -B_2 + iJ = t^A \omega_A = -b^a \omega_a + i v^\alpha \omega_\alpha\ .
\eeq 
Let us now include the R-R forms. Invariance under the orientifold projection 
enforces the expansions 
\beq \label{CCexpansion}
   C_2 = c^a \omega_a \ ,\qquad \qquad  \quad C_4 = \rho_\alpha \tilde \omega^\alpha\ , 
\eeq
where $\omega_a$ was already introduced in \eqref{JBexpansion} and we have 
denoted by $ \tilde \omega^\alpha$ an integral basis of $H^{4}_+(Y,\bbZ)$ dual to $\omega_\alpha$.
Note that in \eqref{CCexpansion} we have only displayed the part of the expansion of $C_4$
which leads to four-dimensional scalars.\footnote{The vectors discussed in the previous 
paragraph arise precisely in the expansion of $C_4$ into appropriate 
three-forms.} Let us now define the even form
\beq \label{def-rho}
  \rho = 1+t^A \omega_A - \cF_{A} \tilde \omega^A + (2 \cF - t^A \cF_A) \epsilon\ ,
\eeq
where $\cF$ is the pre-potential on $\cM_{\rm ks}$ and $\cF_{A}$ is its first derivative
with respect to $t^A$.
The orientifold effective theory including a general pre-potential $\cF$
was derived in refs.~\cite{GL2,TG}.
It was shown there, that the complex coordinates on the K\"ahler manifold $\tilde \cM_{\rm q}$ are obtained in the expansion
\beq \label{def-rhoc}
  \rho_c \equiv  e^{-B_2} \wedge C^{\rm RR} + i \R\big( C \rho \big)= \tau + G^a \omega_a - T_\alpha \tilde \omega^\alpha\ ,
\eeq
where $C^{\rm RR} = C_0 + C_2 + C_4$ and the function $C$ is identified with the dilaton $e^{-\phi}$.
The K\"ahler potential for the complex scalars $\tau,G^a,T_\alpha$ is then 
shown to be 
\bea \label{Kqgeneral}
 K_{\rm q}(\tau,G,T)& =& - 2 \ln \big[i \int_Y \big< C\rho,\overline{C\rho}\big>\big] \\
                    & =& - 2 \ln \big[i|C|^2 \big( 2(\cF -\bar \cF)-(\cF_\alpha +\bar \cF_\alpha)(t^\alpha-\bar t^\alpha) \big) \big]\ , \nn
\eea
where we have inserted the even form $\rho$ defined in \eqref{def-rho} to evaluate the second equality.\footnote{%
\label{wedge-product}The anti-symmetric product between two even forms $\rho,\lambda$ is defined as the alternating wedge product 
$\big<\rho,\lambda \big>= \rho_0 \wedge \lambda_6 - \rho_2 \wedge \lambda_4+\rho_4 \wedge \lambda_2 - \rho_6 \wedge \lambda_0$, where $\rho_p,\lambda_p$ 
are the $p$-form parts of $\rho,\lambda$.}
Note that $K$ is a function of the imaginary part $\I\rho_c= \R(C \rho)$ of $\rho_c$ only. 
This implies that $K$ only depends on the combinations $\tau-\bar \tau$, $G^a-\bar G^a$ and $T_\alpha -\bar T_\alpha$.
For a general pre-potential $\cF$ it is impossible to explicitly write $K$ as the function of $\tau,G^a,T_\alpha$.
This is due to the fact that one would need to express  $\I (C\rho)$ as a function of $\I \rho_c=\R(C \rho)$
 appearing in the $\cN=1$ coordinates \eqref{def-rhoc}. This functional dependence is highly non-polynomial 
 and can only be determined explicitly in specific 
 examples.\footnote{This is equivalent to the problem of solving the attractor equations 
 for $\cN=2$ black holes.}
 Nevertheless, one can derive the K\"ahler metric
 by using the underlying $\cN=2$ special geometry \cite{GL2} or the work of Hitchin \cite{Hitchin} as
 done in \cite{BG}. 
 
 So far we have determined the $\cN=1$ kinetic terms of the scalar and vector 
 fields. Masses for these scalar fields can be 
 generated by a non-trivial superpotential or the presence of D-terms.
 In the rest of the paper we will only discuss the inclusion of a superpotential. 
 In type IIB orientifolds with O3/O7 planes it can be generated by non-vanishing 
 R-R and NS-NS three-form flux $F_3$ and $H_3$ as well as non-perturbative corrections due 
 to D-instantons. It takes the form \cite{Witten,GVW,GKP,KKLT}
 \beq \label{full_super}
    W = \int_Y \Omega(z) \wedge \big(F_3 - \tau H_3 \big) + W_{\text{D-inst}}(\tau,z,G,T,\ldots)\ .
 \eeq
The first term is the well-known Gukov-Vafa-Witten flux superpotential, while the 
second term encodes the D-instanton effects. We will discuss the field dependence 
and modular properties of $W_{\text{D-inst}}$ in section \ref{generalsup}. In order to do that 
it is often convenient to also refer to the underlying F-theory description of 
the orientifold setup. We therefore end this section with some 
remarks on the F-theory embedding and four-dimensional symmetries.

Type IIB orientifolds with O3 and O7 planes arise as 
a special limit of F-theory \cite{Vafa} compactified on particular four-dimensional 
Calabi-Yau manifolds \cite{Sen}. These fourfolds have to admit an elliptic 
fibration 
\beq \label{elliptic}
    Y_4 \rightarrow B_3\ ,
\eeq 
where $B_3$ is some three-dimensional 
base manifold. The complex structure of the torus fiber 
corresponds to the complex dilaton $\tau$ introduced above. 
In general $\tau$ can vary over the base $B_3$.  This implies the 
existence of a modular group $\Gamma_M$ associated to the elliptic fibration. 
This group encodes the monodromies around the singular 
points of the fibration and is a discrete subgroup of the torus 
symmetry group $Sl(2,\bbZ)$. The complete $Sl(2,\bbZ)$ symmetry corresponds to 
the non-perturbative symmetry of type IIB string theory. In the full F-theory compactification 
it is reduced or broken due to the background geometry $Y_4$ \cite{Vafa, BKMT}. Roughly speaking,
the larger the modular group $\Gamma_M \in Sl(2,\bbZ)$, the fewer symmetries survive in the 
effective four-dimensional action.

In this paper we will entirely focus on the orientifold limit reviewed in this section \cite{GKP, FluxReviews}.
It was shown in \cite{Sen} that in this limit the base $B_3$ can 
be obtained as a quotient of a Calabi-Yau manifold by an involution $\sigma$
as discussed above. 
The singularities of elliptic fibration \eqref{elliptic} determine the location of the space-time filling 
O7 planes and D7 branes. However, in the above orientifold limit, both the complex 
dilaton as well as the fields $G^a$ do not vary over the base $B_3$, 
but correspond to chiral fields in four space-time dimensions. In other 
words, in this limit the monodromy group $\Gamma_M$ acts trivially on 
$\tau,G^a$ and we expect that a subgroup $\Gamma_S \subset Sl(2,\bbZ)$ 
survives as a symmetry of the effective action.
This symmetry posses stringent constrains on the $\cN=1$ characteristic 
data of the orientifold compactification in analogy to \cite{FILQ,CFILQ}. In the next sections 
we discuss these conditions in detail. Clearly, a more general 
analysis would consider the full F-theroy compactification and we hope to 
return to this problem in forthcoming work. Let us just remark here, that    
there is no known effective action of twelve-dimensional 
F-theory. The four-dimensional $\cN=1$ effective theory thus has 
to be determined by an M-theory lift. More precisely, one compactifies 
M-theory on the elliptically fibered fourfold $Y_4$ to obtain a three-dimensional 
effective theory. This theory is then lifted to four-dimensions by growing 
an extra non-compact dimension. The F-theory moduli thus arise from the 
expansion of the M-theory fields, such as the three-form $C_M$, 
into harmonics of $Y_4$. A detailed discussion of the derivation 
of the effective action can be found, for example, in refs.~\cite{Mayr:1996sh,HL,TG}.

\subsection{Perturbative and non-perturbative $\alpha'$ corrections in the orientifold large volume limit \label{largevolume}}

In this section we simplify the discussion and work in the large volume limit of the orientifold $Y/\sigma$. 
This implies that we consider the regime where $v^\alpha$ is large.
Note that this is not the same as demanding that all $v^A$ are large on the underlying Calabi-Yau
manifold, since $v^a=0$ in the orientifold setup.
In other words, the contributions depending on $t^a = -b^a$ are not necessarily
suppressed in the large volume limit of the orientifold. We therefore include 
the non-pertubative $\alpha'$ corrections inherited from the underlying $\cN=2$ theory.
More precisely, we obtain in this limit a 
pre-potential of the form\footnote{%
Note that in general 
$\cF$ can also admit a cubic and linear term of the form $B_{AB} t^A t^B$,  $A_A t^A$.
However, since $A_A,B_{AB}$ are always real it is easy to check 
that they do not appear in the K\"ahler potential \eqref{Kqgeneral}. They only 
correct the coordinates $T_\alpha$ and we will not consider these contributions 
in the following.}  
\bea \label{simple_F}
  \cF &=& \cF_{\rm class}+ \cF_{\rm pert}+\cF_{\rm b}\\
  &=&  - \tfrac{1}{3!} \cK_{ABC} t^A t^B t^C  - \tfrac{i}{2}\zeta(3)\chi + i \sum_{\beta \in H_2^-(Y,\bbZ)} n_{\beta}^0 \ \text{Li}_3 (e^{i k_a t^a}) \ , \nn
\eea
where $k_a = \int_\beta \omega_a$ with $\omega_a$ being an integral basis of $H^2_-(Y,\bbZ)$.
Let us discuss the three contributions in \eqref{simple_F} in turn.
 The cubic term $\cF_{\rm class}$ corresponds to the classical contribution and
we denote the triple intersections of the integral basis $\omega_A \in H^{2}(Y,\bbZ)$ by 
\beq \label{triple_inters}
  \cK_{ABC} = \int_Y \omega_A \wedge \omega_B \wedge \omega_C\ .
\eeq
Note that in the orientifold setup consistency requires that for the spilt $\omega_A=(\omega_\alpha,\omega_a)$
the following intersections have to vanish: 
\beq\label{inter_constraints}
   \cK_{\alpha \beta a}=\cK_{abc} = 0\ .
\eeq
In other words only the intersections $\cK_{\alpha \beta \gamma}$ and $\cK_{\alpha ab}$ with zero or two negative 
indices can appear in \eqref{simple_F}. 
The second term $\cF_{\rm pert}$ in \eqref{simple_F} is proportional to the Euler characteristic $\chi = 2(h^{(1,1)}-h^{(2,1)})$ of $Y$.
It corresponds to an $(\alpha')^3$ perturbative correction of the effective action and was first considered 
in orientifold setups in ref.~\cite{BBHL}. 

The third term $\cF_{\rm b}$ is inherited from 
the non-perturbative  $\alpha'$ corrections of the $\cN=2$ pre-potential and was not discussed in the literature 
so far. In the large volume limit 
of the orientifold  only the terms depending on the B-field moduli $t^a=-b^a$
 survive in the third polylogarithm Li$_3(x)=\sum_{n>0} n^{-3} x^n$.
All other contributions are suppressed exponentially by the volume of the curves in $H_2^+(Y,\bbZ)$.
In other words, only the terms proportional to  the integer genus zero Gopakumar-Vafa invariants $n^0_{\beta}$ \cite{GV}
for a curve $\beta$ in the negative eigenspace $H_2^-(Y,\bbZ)$ remain in the pre-potential.
They can be determined for many explicit examples of Calabi-Yau manifolds my using 
mirror symmetry \cite{Mirrorbook}. However, we have
to make a cautionary remark on the convergence of the expansion \eqref{simple_F}.  Since the polylogarithm 
$ \text{Li}_3 (e^{i k_a t^a}) $ is bounded $\cF_{\rm b}$ appears divergent when summing over all $\beta$. 
This would be very generically the case if $\beta$ is not restricted to any sublattice in $H_2(Y,\bbZ)$ since 
the Gopakumar-Vafa invariants grow very rapidly. However, in the expression \eqref{simple_F} for $\cF_{\rm b}$
we only sum over degrees $k_A$ which are of the form $k_A=(0,k_a)$, i.e.~vanish on the positive 
eigenspace of the orientifold. There are indeed examples for which the $n_{\beta}^0$ 
truncates on such a sublattice $(0,k_a)$.\footnote{We are grateful to A.~Klemm for discussions on this 
point.} More generally, in case $\cF_{\rm b}$ is not finite this can be traced back to 
the fact that we are actually working in the wrong coordinates $t^a$. Before restricting 
to the orientifold limit $\I t^a\rightarrow 0$ the expression $\cF_{\rm b}$ has to be resummed in terms 
of dual coordinates valid around $\I t^a=0$. One is then able to implement the orientifold 
projection with a finite $\cF_{\rm b}$.  In the following we will simply assume that $\cF_{\rm b}$ is finite 
when restricting our general considerations to appropriate specific examples.

In order to determine the $\cN=1$ coordinates 
we first insert the large volume pre-potential \eqref{simple_F} into the 
definition \eqref{def-rho} of the even form $\rho$. Due to the presence 
of the $\alpha'$ corrections $\cF_{\rm pert} + \cF_{\rm b}$ the classical 
expression $\rho_{\rm class} = e^{-B_2+iJ}$ will receive non-trivial corrections. 
However, it is easy to check that these corrections will not contribute 
to the definition of the $\cN=1$ coordinates $\tau,G^a,T_\alpha$
defined in \eqref{def-rhoc}.
A straightforward computation shows
that $\tau,G^a,T_\alpha$ are given in 
terms of the real coordinates introduced in \eqref{JBexpansion} and \eqref{CCexpansion} 
by
\bea  \label{def-tauG}
  \tau  &=& C_0 + ie^{-\phi}\ , \qquad \qquad  G^a = c^a - \tau b^a\ ,\\
   \label{def-T}
    T_\alpha &=&\tfrac{1}{2} ie^{-\phi} \cK_{\alpha \beta \gamma} v^\beta v^\gamma - \tilde \rho_\alpha-\frac{1}{2(\tau -\bar \tau)} \cK_{\alpha ab} G^a (G-\bar G)^b\ ,
\eea
where $\tilde \rho_\alpha = \rho_\alpha - \frac12 \cK_{\alpha ab} c^a b^b$.
These are precisely the coordinates introduced in ref.~\cite{GL1}.\footnote{In contrast to ref.~\cite{GL1} we rescaled the 
coordinates $T_\alpha =  \frac{2i}{3} T^{\rm ref.}_\alpha$ and identified $\tilde \rho_{\alpha} = \rho_\alpha^{\rm ref.}$.}
However, in contrast to the classical results the K\"ahler potential $K_{\rm q}$ is now corrected by the $\alpha'$
contributions encoded by $\cF_{\rm pert}+\cF_{\rm b}$ in \eqref{simple_F}.

Let us make this more precise and evaluate the K\"ahler potential for the large volume pre-potential \eqref{simple_F}.
Inserting $\cF$ into the general expression \eqref{Kqgeneral} for $K_{\rm q}$ one derives
\beq
   K_{\rm q} =- 2\ln \Big[e^{-2\phi} \big(\tfrac{1}{3!}\cK_{\alpha \beta \gamma} v^\alpha v^\beta v^\gamma + 2\zeta(3) \chi - 4 \I \cF_{\text{b}} \big) \Big]\ .
\eeq
In this expression the non-perturbative corrections inherited from the underlying 
$\cN=2$ theory take the form  
\bea
 \I \cF_{\text{b} }(\tau,G) &=& \tfrac{1}{2}\sum_{\beta\in H_2^-(Y,\bbZ)} n^0_{\beta} \, \Big[ \text{Li}_3 \Big(e^{ i \frac{k_a (G^a-\bar G^a )}{\tau-\bar \tau}} \Big) + \text{Li}_3 \Big(e^{ -i \frac{k_a (G^a-\bar G^a )}{\tau-\bar \tau}} \Big)\Big]\ , \nn \\
 \label{ImFb}
 &=& \sum_{\beta\in H_2^-(Y,\bbZ)}  \sum^{\infty}_{n=1}\  \frac{n^0_{\beta}}{n^3}\ \cos \left( n \frac{k_a (G^a-\bar G^a )}{\tau-\bar \tau}\right)\ ,  
\eea
where $k_a = \int_\beta \omega_a$ as in \eqref{simple_F}.
This implies that the moduli dependence on $\tau,G^a$ of both $\alpha'$ corrections to the K\"ahler potential can
be determined explicitly.
Rescaling the K\"ahler deformations  $v^\alpha$ to the Einstein frame we can write 
$K_{\rm q}$ into the form 
\beq
   \label{Kqgen}
   K_{\rm q}= - \ln\big[ -i (\tau -\bar \tau) \big]- 2 \ln\Big[ V_{E} + 
   \tfrac{1}{(2i)^{3/2}} \big(\tau -\bar \tau \big)^{3/2}  \big[2\zeta(3) \chi -4 \I \cF_{\text{b}}\big] \Big]\ ,
\eeq
where $ V_{E}(\tau,G,T)$ is the Einstein frame volume of the Calabi-Yau orientifold and $\cF_{\rm b}(\tau,G)$
is explicitly given in \eqref{ImFb}.
The large volume 
K\"ahler potential \eqref{Kqgen} includes the special cases derived in refs.~\cite{GKP,BBHL,GL1}.
Here we were able to include the non-perturbative contribution $\cF_{\rm b}(\tau,G)$
and have shown that they can be expressed as explicit functions in $G^a-\bar G^a$ and $\tau-\bar \tau$.
In the next section we will discuss the invariance of the general K\"ahler potential 
\eqref{Kqgen} under the $Sl(2,\bbZ)$ symmetry 
of type IIB string theory as well as shifts in the B-field.

\subsection{Symmetries of the K\"ahler potential \label{Kaehlersymm}}

In this section we discuss the transformation properties of the K\"ahler potential under 
dualities inherited from the ten-dimensional type IIB string theory. We will focus on the 
$Sl(2,\bbZ)$ symmetry of type IIB as well as shifts in the NS-NS two-form $B_2$.

Let us begin by discussing the symmetry of $K$ under shifts of  the 
NS-NS two-form $B_2$. More precisely, we will consider 
\beq \label{B2shift}
  B_2 \quad \rightarrow \quad B_2 + 2\pi \chi_2\ , \qquad \quad \chi_2=n^a \omega_a\ ,
\eeq
where $\chi_2$ is an integral two form in $H^{2}_-(Y_E,\bbZ)$. For this transformation 
we easily verify that the K\"ahler potential is invariant. The Einstein frame 
volume $V_E$ in \eqref{Kqgen} is invariant due to its purely geometrical origin, while 
the perturbative contribution from $\cF_{\rm pert}$ is independent of $B_2$ and 
hence trivially invariant. Only the non-perturbative corrections encoded by $\cF_{\rm b}$
explicitly depend on $B_2$. However, $B_2$ only arises through the exponential 
$\exp(-i\int_\beta B_2)$ which is invariant under integral shifts. We thus conclude that $K$ is indeed invariant 
under \eqref{B2shift}.
In contrast, we will see in the next section that the $\cN=1$ coordinates $G^a,T_\alpha$
transform non-trivially under the shifts \eqref{B2shift}. This will allow us to infer valuable 
information about the moduli dependence of the D-instanton superpotential in \eqref{full_super}.

Let us turn to the symmetry inherited from the underlying type IIB theory.
Recall that type IIB string theory admits the discrete 
symmetry group  $Sl(2,\bbZ)$. Denoting the ten-dimensional dilaton-axion  
as $\tau= C_0 + i e^{-\phi}$ this group acts by modular transformations and rotates 
the ten-dimensional NS-NS and R-R two-forms $B_2$ and $C_2$ into each other. 
More explicitly, we have 
\beq \label{modular-tau}
     \tau \quad \rightarrow\quad \frac{a\tau+b}{c\tau +d} \ ,\qquad \qquad \left(\begin{array}{c} C_2\\ B_2\end{array} \right) \quad \rightarrow\quad \left(\begin{array}{c}a\,C_2+b\, B_2\\c\,C_2+d\, B_2 \\\end{array} \right)\ ,
\eeq
where the integer matrix {\footnotesize $\left(\begin{array}{cc}a&b\\ c& d \end{array}\right)$} is an element of $Sl(2,\bbZ)$.%
\footnote{
Here we have been a bit sloppy with factors of $2\pi$, which however can be restored easily.}
These transformations include in particular the map $\tau \rightarrow -1/\tau$ which 
inverts the string coupling and corresponds to the strong-weak duality known as S-duality.
Compactifying type IIB string theory on a Calabi-Yau orientifold background
can reduce the symmetry group $Sl(2,\bbZ)$ to a subgroup $\Gamma_S$
as discussed at the end of section \ref{orirev}.

Let us now check how the K\"ahler potential and K\"ahler coordinates 
transforms under modular transformations \eqref{modular-tau} in $\Gamma_S$.
We concentrate in the following on the large volume compactification characterized 
by the $\alpha'$ corrected pre-potential \eqref{simple_F}.
Using the explicit expressions \eqref{def-tauG} and \eqref{def-T} for $G^a, T_\alpha$ we note that these $\cN=1$
coordinates  
transform under \eqref{modular-tau} as %
\footnote{For $T_\alpha$ to transform as in \eqref{tautransgen} we have used that $e^{-\phi/2} v^\alpha$ and 
$\tilde \rho_\alpha$ are invariant under \eqref{modular-tau}. The combination $e^{-\phi/2} v^\alpha$ is precisely the invariant Einstein 
frame K\"ahler structure deformation, while $\tilde \rho_\alpha$ arises in the expansion of an $Sl(2,\bbZ)$ invariant $\tilde C_4$ with 
field strength $F_5 = d \tilde C_4 -\frac12 dB_2 \wedge C_2 + \frac12 B_2 \wedge dC_2$. We have also 
used that $(\tau -\bar \tau)^{-1} \rightarrow  (c\tau +d)^2 (\tau -\bar \tau)^{-1} -c(c\tau+d)$.}
\beq \label{tautransgen}
 G^a \ \rightarrow \ \frac{G^a}{c \tau + d}\ ,
  \qquad \qquad T_\alpha\ \rightarrow\ T_\alpha + \frac{1}{2} \frac{c\ \cK_{\alpha a b} G^{a} G^b}{c\tau+d}\ .
\eeq
where $a,b,c,d$ are the entries of an element of $ \Gamma_S$.
We next analyze how the perturbatively corrected K\"ahler potential \eqref{Kqgen} transforms under \eqref{tautransgen}. 
It is very easy to evaluate the transformation properties of the first term in \eqref{Kqgen} since
\beq
   (\tau -\bar \tau)^{-1} \quad \rightarrow \quad |c\tau +d|^2 (\tau -\bar \tau)^{-1}\ .
\eeq
We thus have to focus on the transformation of the combination 
\beq \label{pertcorr}
 V_{E}(\tau,G,T) + \tfrac{1}{(2i)^{3/2}} \big(\tau -\bar \tau \big)^{3/2}  \big[2\zeta(3) \chi -4 \I \cF_{\text{b}}(\tau,G)\big] \ .
\eeq
Clearly, the Einstein-frame volume $V_E$ is invariant under $\Gamma_S$, since it is 
a purely geometric quantity. Note however, that invariance does not hold for the $\alpha'$
correction in \eqref{pertcorr}. This can be traced back to the fact that we did not include all 
corrections relevant in this large volume limit. Analogously to the discussion in refs.~\cite{GG,RRSTV} 
one can argue that also corrections due to D(-1) branes as well as the reduction of D1
instantons have to be included. These couple to 
the complex dilaton $\tau$ and $G^a$ and can complete the $\alpha'$ correction in \eqref{pertcorr} 
in a modular invariant form. 
We propose that by including these contributions the large volume K\"ahler potential $K_{\rm q}$ takes the form
\beq
   \label{Kqgenmod}
   K_q= - \ln\big[ -i (\tau -\bar \tau) \big]- 2 \ln\Big[ V_{E} +  \tfrac12 \chi\, f(\tau,\bar \tau) -4 g(\tau,\bar \tau,G,\bar G)\Big]\ ,
\eeq
and transforms under $\Gamma_S$ as 
\beq \label{transK}
   e^{K} \quad \rightarrow \quad |c\tau +d|^2 e^K\ .
\eeq
In general it is hard to determine the precise form of the modular invariant forms 
$f(\tau,\bar \tau)$ and $g(\tau,\bar \tau,G,\bar G)$. In the remainder of this section
we will discuss some properties of $g,f$ as well as some candidate modular 
completions. A calculation of 
$f,g$ might be possible by restricting the class of Calabi-Yau manifolds to $K3$ fibrations 
where heterotic-F-theroy duality can be applied.

In the following we will first discuss the modular invariant function $f(\tau,\bar \tau)$
in \eqref{Kqgen}. In order to do that, we recall that in ref.~\cite{GG} a similar problem 
arose  in the computation of the $R^4$ correction to the
ten-dimensional type IIB supergravity action. In this ten-dimensional setup, 
an additional analysis of the properties of the $\tau$-dependent coefficient 
$\hat f(\tau,\bar \tau)$ led to the identification 
\beq \label{sumhatf}
  \hat f(\tau,\bar \tau) = \sum_{(n,m) \in P} \frac{(\tau-\bar \tau)^{3/2}}{(2i)^{3/2}|m+n \tau|^3} \ ,
\eeq
where $P=\bbZ^2/(0,0)$ is a two-dimensional lattice without the origin.
This non-holomorphic Eisenstein series includes indeed the perturbative correction 
in \eqref{pertcorr}, when $n=0$ in the sum \eqref{sumhatf}. Moreover, it is invariant under the 
full group $Sl(2,\bbZ)$ and hence a candidate modular completion of the K\"ahler 
potential. It was also conjectured in ref.~\cite{RRSTV} that the function \eqref{sumhatf}
is the correct modular completion of the analog situation in the underlying $\cN=2$
theory. In our setup one might want to restrict the sum in \eqref{sumhatf} only to orbits of 
the subgroup $\Gamma_S$. However, in any case modularity together with the limit $n=0$ 
alone seems not sufficient to fix the form of 
$f(\tau,\bar \tau)$ in \eqref{Kqgen}. Additional conditions such as the singularity 
structure or the suppression of further mixed contribution 
are needed to determine $f(\tau,\bar \tau)$ unambiguously. 
This is in general hard and beyond the scope of this paper. For the general discussion 
of the superpotential we will simply assume that such a modular completion exists, while 
for our explicit example in section \ref{EnriquesO} we will find that $\chi=0$.

Let us also briefly discuss the modular completion $g(\tau,\bar \tau,G,\bar G)$ 
of the non-perturbative $\alpha'$ corrections inherited from $\cN=2$.
The corrections we are missing in our computation are the D1 branes 
dual to the world-sheets inducing the contribution $\cF_{\rm b}$. More precisely, 
we need to include the whole set of $(p,q)$ strings \cite{Schwarz:1995dk,Witten:1995im} to restore $\Gamma_S$ duality.
Again we are facing the problem that such corrections are hard to compute in general 
and we can only discuss some candidate solution for $g$.
In ref.~\cite{RRSTV} the modular completion of the underlying $\cN=2$ 
quaternionic geometry was conjectured to arise from 
a summation over all $Sl(2,\bbZ)$ images of the world-sheet instanton
corrections. In the orientifold limit this leads to the following definition 
of a modular invariant $\hat g$
\beq
  \hat g(\tau,\bar \tau,G,\bar G) = \sum_\beta n_{k_a} \sum_{(m,n)\in P}  \frac{(\tau-\bar \tau)^{3/2}}{(2i)^{3/2}|n+m \tau|^3} \cos \Big((n+\tau m)  \frac{k_a(G^a - \bar G^a)}{\tau-\bar \tau} -  m k_a G^a \Big)\ .
\eeq
This sum encodes all images under $Sl(2,\bbZ)$ of the world-sheet instanton corrections in $\I \cF_{\rm b}$
divided by stabilizer group generated by shifts $\tau \rightarrow \tau +1$. In general one might also want 
to restrict to orbits of the subgroup $\Gamma_S$.
It is not hard to check that $\hat g$ contains the contribution $\I \cF_{\rm b}$ for $m=0$.
Once again we have to remark that even though $\hat g$ has the desired 
properties, the true correction $g$ is expected to be more complicated. 
It would thus be desirable to find independent ways to calculate 
$g$ for specific setups. In the example of section \ref{EnriquesO} all non-perturbative 
$\alpha'$ corrections will be absent such that no $g$ is inherited from $\cN=2$.

Before moving on to the discussion of the superpotential, 
let us compare the question of determining $f(\tau,\bar \tau)$ 
and $g(\tau,\bar \tau,G,\bar G)$  to a somewhat similar situation within 
topological string theory on a Calabi-Yau threefold \cite{BCOV,YY, Klemm, GKMW}. The symmetry 
group in this case is the target space duality group arising from the 
monodromies around singularities in the moduli space. One 
can thus attempt to parametrize the non-perturbative 
corrections by modular forms of this duality group which form a finite ring. 
Fortunately, the singularity structure for the topological string partition 
function is often known and additional boundary conditions allow to 
fix the precise modular forms encoding the non-perturbative corrections
at least up to a certain genus. These boundary conditions arise from the 
singularities of the moduli space or through the application of 
string-string dualities (see e.g.~\cite{Klemm,GKMW}).
One might thus hope that to redo a similar analysis 
in the $\cN=1$ theories discussed in this work. Clearly, one of the obstacles 
is the non-holomorphicity of the K\"ahler potential as well as the presence 
of additional perturbative corrections. For the holomorphic 
$\cN=1$ superpotential  this situation is improved as we will discuss in the 
next section.

\subsection{D-instanton superpotentials in type IIB orientifolds \label{generalsup}}

Let us now discuss the D-instanton superpotential arising in type 
IIB orientifolds with O3/O7 planes. The instantons contributing 
to the superpotential are typically Euclidean D3 branes wrapped 
around special four-cycles inside the Calabi-Yau orientifold. 
In order to give the precise conditions
when such a potential arises, one has to embed this orientifold setup into 
an F-theroy compactification.  These conditions have been investigated
first by Witten in \cite{Witten} and later refined for compactifications with 
background fluxes \cite{zeromodes}. Since here our primary interest is the 
definition of a symmetry invariant superpotential for a generic orientifold 
compactification, we will directly go to the orientifold and assume that 
these conditions are satisfied for the cycles under consideration. 

In the type IIB orientifolds discussed in the previous sections the instanton 
superpotential arises from specific Euclidean D3 branes. Let us consider such 
a brane warp around a devisor $\Sigma$ in $Y/\sigma$.  
We will pick the devisor such that it non-trivially contributes to 
the superpotential. 
Schematically these contributions are of the form 
\beq
    f(X^I)\ e^{-V_\Sigma + i \phi_\Sigma}
\eeq
where  $V_{\Sigma}$ is the Einstein-frame volume of $\Sigma$ and $\phi_\Sigma$ is the integral 
of the R-R four-form $C_4$ over $\Sigma$. The function $f(X^I)$ can depend on 
other chiral multiplets in the spectrum and we will be the main focus of our 
considerations. Before turning to the discussion of $f$, let us
first note that the form of the exponential is not yet exact, since 
we are missing the coupling to the lower R-R forms and the B-field in the exponential.
Recall that the effective 
action on the word-volume of the Euclidean D3 brane takes the 
form  
\beq
   S^{D3}=iT_{D3}  \int_{\cW_4} d^4 \lambda e^{-\phi} \sqrt{\det\big(g-B_2+F\big)}+ T_{D3} \int_{\cW_4} C^{\text{RR}} \wedge e^{-B_2 + F} \ ,
\eeq
where $C^{\text{RR}}=C_0 + C_2 + C_4$ are the Ramond-Ramond fields and $F$ is the fieldstrength on the 
brane. The first and second term correspond to the Born-Infeld and Chern-Simons 
coupling respectively. In order that the D-instanton preserves 
supersymmetry it has to wrap a supersymmetric cycle. 
Applying the standard calibration conditions for supersymmetric branes 
we find that the correct couplings to the R-R forms and the B-field \cite{BBS}. 
The correct superpotential contribution is thus proportional to
\beq \label{exponent}
   \text{exp}\Big[ -  \tfrac12 \int_\Sigma  e^{-\phi} \big(J \wedge J- B_2 \wedge B_2\big) - i \int_\Sigma \big( C_4 - C_2\wedge B_2 +\tfrac12 C_0 B_2\wedge B_2\big) \Big]\ .
\eeq
Note that the first term under the first integral is $V_{\Sigma}$, since the K\"ahler form $J$ is evaluated in 
the string-frame metric. The expression \eqref{exponent} is precisely exp$(-i\int \rho_c)$ with $\rho_c$ introduced 
in \eqref{def-rhoc}. Thus we find that the generic superpotential is of the expected form 
\beq \label{Dinst_superpot}
   W_{\text{D-inst}} = \sum_\Sigma f_\Sigma(X^I) \ e^{i n^{\ \alpha}_\Sigma\, T_\alpha}\ ,\qquad \qquad n_{\Sigma}^{\ \alpha} = \int_\Sigma \tilde \omega^\alpha\ ,  
\eeq
where $n_\Sigma^{\ \alpha}$ are integers for $\Sigma \in H_{4}(Y,\bbZ)$ and $\tilde \omega^\alpha \in H^{4}_+ (Y,\bbZ)$.
We are now in the position to discuss the moduli dependence  of $f(X)$ in more detail.

So far we did not discuss the holomorphic function $f(X)$. In general, it can 
depend on various other moduli $\{X^I\}$ of the orientifold or underlying F-theory compactification.
As in \eqref{elliptic} we denote the elliptically fibered fourfold corresponding to the orientifold by $Y_4$. 
The moduli dependence of $f$ can arise from:
\begin{itemize}
\item[(a)] the complex structure deformations of $Y_4$: in the orientifold limit these 
include the complex dilaton $\tau$ corresponding 
  to the complex structure of the elliptic fiber, the complex structure deformations of $Y/\simga$
  as well as the D7 brane moduli,
\item[(b)] the $h^{(2,1)}$ complex scalars arising in the expansion of $C_M$ in $H^{(2,1)}(Y_4)$: these 
include the complex scalars $G^a$ as well as Wilson lines of the D7 brane,
\item[(c)] the complex coordinates $x^i$ labeling the position of space-time filling D3-branes in $Y_4$ or $Y/\simga$.
\end{itemize}
In the following we will discuss $f(X)$ as a function of the complex dilaton $\tau$, 
the moduli $G^a$ arising by expanding the type IIB NS-NS and R-R two-form. 
An analysis of the dependence of $f(X)$
on the positions of the space-time filling D3-branes $x^i$ on $Y/\simga$
can be found in \cite{Ganor,Baumann:2006th}.

It turns out that a direct computation of the function $f(X)$ is in 
general very hard and involves the evaluation of appropriate determinants \cite{Witten}.
However, we can already learn much about $f$ by studying the transformation 
properties of the superpotential and the K\"ahler potential under 
shifts and modular transformations. This was already initiated in refs.~\cite{Witten:1996hc,Ganor} for 
M- and F-theory compactifications were only a local analysis can be performed. 
Here we will make this discussion very concrete for the type IIB orientifolds 
studied in section \ref{orirev} and focus on its dependence on $\tau,G^a$
decomposing $f(X) = A_0 \Theta(\tau,G^a)$, with $A_0$ depending on the remaining
moduli. We thus write 
\beq \label{superpotexp}
   W_{\text{ D-inst}} = A_0 \sum_{\Sigma} \Theta_{\Sigma}(\tau,G) e^{in_{\Sigma}^{\ \alpha}\, T_\alpha }\ .
\eeq

Let us now investigate the transformation properties of the coefficients $\Theta_{\Sigma}(\tau,G^a)$ in more detail.
We will first discuss the duality transformations induced by modular changes 
of the complex dilaton $\tau$ as given in \eqref{modular-tau}. In section \ref{Kaehlersymm} we have argued that 
$e^K$ transforms as given in equation \eqref{transK} under modular transformations.
From this we conclude that the superpotential has to change as \footnote{%
In the following we will not include a possible phase. For a related discussion of the possibility 
to include such a phase factor see, for example, refs.~\cite{FILQ, CFILQ}.} 
\beq \label{transWundertau}
   W  \quad \rightarrow \quad (c\tau +d)^{-1} W \ .
\eeq
To see this, we note that the combination $e^{K}|W|^2$ has to be invariant since 
it determines, for example, in the physical gravitino mass.
Equation \eqref{transWundertau} exactly states 
that $W$ has to be a modular form of weight $-1$ under the duality group $\Gamma_S$.
Let us note that this is obviously true for the flux superpotential $W= \int \Omega\wedge (F_3 -\tau H_3)$ in \eqref{full_super}.
For the D-instanton superpotential \eqref{superpotexp} we will see momentarily, that this imposes constraints
on the functions $\Theta_{\Sigma}(\tau,G)$.

The second transformation we will consider are the shifts \eqref{B2shift}
in the NS-NS two-form $B_2$. More precisely,
let us transform the orientifold coordinates by $b^a \rightarrow b^a + 2\pi n^a$.
From the definitions \eqref{def-tauG} and \eqref{def-T} of the 
coordinates $G^a,T_\alpha$ we deduce that 
\bea \label{Gshift}
  G^a  \quad &\rightarrow& \quad G^a - 2 \pi \tau n^a \ ,\\
  T_\alpha \quad&\rightarrow& \quad T_\alpha - 2\pi \cK_{\alpha ab} n^a G^b + 2 \pi^2  \tau  \cK_{\alpha ab} n^a n^b \nn \ .
\eea
As we have seen in section \ref{Kaehlersymm}, it is not hard to check that this is a symmetry 
of the orientifold K\"ahler potential. Due to the invariance of the combination $e^K |W|^2$ we conclude that
$W$ can only transform by a trivial phase factor and is otherwise invariant. 
Invariance of $W$ together with the fact that $T_\alpha$ transforms as in \eqref{Gshift}
restricts the coefficient functions $\Theta_{\Sigma}(\tau,G)$
of the instanton superpotential \eqref{superpotexp} as we will discuss next.

We can now infer the properties of the functions $\Theta_{\Sigma}(\tau,G)$ 
appearing in \eqref{superpotexp}. Our strategy is to use the fact that $W$ 
is a modular form of weight $-1$ but otherwise invariant under \eqref{modular-tau}, \eqref{tautransgen} and \eqref{Gshift}.
Since $e^{i n_\Sigma^{\ \alpha} T_\alpha}$ in \eqref{superpotexp} transforms non-trivially under these symmetries also $\Theta_{\Sigma}(\tau,G)$
has to transform in order to ensure the correct modular properties of $W$.
It turns out that the $\Theta$'s are 
generalizations of the well-known theta functions, or more precisely
appropriate holomorphic Jacobi forms.\footnote{Holomorphicity here only means that
$\Theta_\Sigma(\tau,G)$ is independent of $\bar \tau,\bar G^a$ and 
does not restrict the singularity structure.} To summarize their properties 
we simplify our analysis and restrict our attention 
to the case where only one $T\equiv T_{\alpha'}$ transforms non-trivially
under the above groups. In other words, we will 
assume here that the only non-vanishing intersection with negative 
indices is $\cK_{\alpha' a b}=-C_{ab}$. 
We also denote $n^{\ \alpha'}_{\Sigma}=n$. The Jacobi form $\Theta_n(\tau,G)$
then turns out to be of weight $-1$ and index $n$. In other words, 
under the transformation \eqref{tautransgen}
this form transforms as
\beq \label{transtheta1}
  \Theta_n(\tau,G) \ \rightarrow \ (c\tau+d)^{-1} \text{exp} \Big({\frac{n i}{2} \frac{c\ C_{a b} G^{a} G^b}{c\tau+d}}\Big)\ \Theta_n (\tau,G) \ ,
\eeq
which is consistent with the required transformation behavior \eqref{transWundertau}.
Also the transformation \eqref{Gshift} of $e^{inT}$ is cancelled by the corresponding Jacobi form 
$\Theta_n$ since 
\beq \label{transtheta2}
    \Theta_n(\tau,G) \ \rightarrow \ \text{exp} \big(- 2\pi i n C_{ab} n^a G^b + 2 \pi^2 i n \tau  C_{ab} n^a n^b \big)\ \Theta_n (\tau,G)
\eeq
under the transformation \eqref{Gshift}. Carefully restoring factors of $2\pi$
the transformations \eqref{transtheta1} and \eqref{transtheta2} are 
exactly the transformation properties of 
Jacobi forms. For only one 
field $G^a$, the theory of Jacobi forms is 
extensively reviewed by Eichler and Zagier in ref.~\cite{EZ}. The more general situation including vectors 
$G^a$ is discussed, for example, in the work of Borcherds \cite{binfinite} (section 3). 

Before turning to the example in the next section,
let us summarize some classical results about candidate Jacobi forms $\Theta_n$ \cite{EZ,binfinite}. In order to do that 
we introduce the theta functions of weight $s/2$ and index $m$ by setting 
\beq
  \theta_{(m)\ L + r}(\tau, G) = \sum_{n_a \in L + r} e^{i \tau n^2/2} e^{m i G^a n_a}\ ,\qquad n^2=C^{ab} n_a n_b\ ,
\eeq
where $L$ is some positive definite rational lattice of dimension $s$, and 
$r$ is some vector which admits an expansion in a basis of $L$ 
with rational coefficients.
It can be shown that any Jacobi form $\Theta_n$ can be written as
a sum of products of the theta functions $ \theta_{(m)\ L + r}$ and 
modular forms $\tilde \eta(\tau)$. Heuristically, we can write
\beq \label{def-Thetan}
  \Theta_n(\tau,G) = \sum  \frac{\theta_{(n)}(\tau, G)}{\tilde \eta(\tau)}\ .
\eeq
This form is well known from various other perspectives. For example, it was shown 
in \cite{Alvarez-Gaume:1987vm} that the partition function of a chiral boson on a genus one surface is of this form. 
More importantly, also the partition function 
of the M5 brane takes a form similar to \eqref{def-Thetan} as was first discussed in ref.~\cite{Witten:1996hc}.  
This is no surprise, since we know that the F-theory lift of the D3 instantons are six-dimensional 
branes. Analyzing F-theory from the M-theory point of view as mentioned at the end of 
section \ref{orirev} these six-dimensional branes are M5 branes wrapped around four-cycles 
in the base $B_3$ of \eqref{elliptic} as well as on the two-dimensional fiber. 

Clearly, an important task is to explicitly find the correct Jacobi forms $\Theta_n(\tau,G)$ for 
specific examples. One suspects that this problem is more tractable then determining the 
modular corrections to the K\"ahler potential due to the holomorphicity of $W$ and the absence 
of perturbative corrections. Ideally, one likes to use physical arguments, for example 
on the singularity structure of $W$, to restrict the set of candidate Jacobi forms to a finite set.
Computing $W$ in a particular limit, e.g.~an orbifold limit, might then determine the 
correct linear combination to appear in the full $W$.
In the next section, we will take a different route in the study of the Enriques orientifold. 
We will use some intuition from the topological strings on the 
Enriques Calabi-Yau to propose a candidate $W$ including non-trivial 
Jacobi forms $\Theta_n$.

\section{D-instantons and the Enriques orientifold \label{EnriquesO}}

In this section we discuss one type IIB orientifold compactification in
more detail and illustrate some of the general story outlined in the previous section. 
We construct an orientifold of the Enriques Calabi-Yau $Y_E$ and argue that
the quantum corrections are under particular control. It is also 
shown how the $\cN=1$ K\"ahler manifold $\tilde \cM_{\rm q}$ inside the $\cN=2$ quaternionic 
space can be identified with the original special K\"ahler moduli 
space times a $Sl(2,\bbR)/U(1)$ factor. In this duality the new complex coordinates contain the 
R-R fields as in \eqref{def-rhoc} and provide the correct couplings to 
D-instantons. We use this identification to translate instanton 
expansions known from topological string theory
on $Y_E$ to the corresponding physical orientifold setup.
This leads us to propose a specific D-instanton superpotential 
for the Enriques orientifold.

\subsection{Enriques Calabi-Yau and counting of D(-1)-D1-D3 states \label{GeometryN=2}}

Let us begin by reviewing some basic facts about the Enriques Calabi-Yau $Y_E$
and its moduli space. The Enriques Calabi-Yau takes the form $Y_E = (K3 \times \mathbb{T}^2)/\bbZ_2$, 
where the $\bbZ_2$ acts as an inversion of the complex coordinate of $\mathbb{T}^2$ and as the Enriques involution 
on $K3$ \cite{Borcea,FHSV, SurfaceB}. $Y_E$ has holonomy group $SU(2) \times \bbZ_2$. This implies that type
II string theory compactified on the Enriques Calabi-Yau will lead to a four-dimensional
theory with $\cN=2$ supersymmetry. Nevertheless, due to the fact that it does not have the
full $SU(3)$ holonomy of generic Calabi-Yau threefolds, various special properties of
$\cN=4$ compactifications on $K3\times \mathbb{T}^2$  are inherited.

In order to discuss the moduli space of $Y_E$ we first need 
to summarize the cohomology on this Calabi-Yau manifold.
We review in appendix \ref{EnriquesGeom} that the two-form and 
three-from integral cohomologies can be identified 
with the following lattices \cite{FHSV}
\bea \label{EnriquesCohomology2}
   H^{2}(Y_E,\bbZ) &\cong& \bbZ \oplus \Gamma^{1,1} \oplus \Gamma_{E_8}(-1)\ ,\\
  \label{EnriquesCohomology3}
   H^{3}(Y_E,\bbZ) &\cong& \big( \Gamma^{1,1} \oplus \Gamma_{E_8}(-1) \oplus \Gamma^{1,1}_g \big)\oplus \big( \Gamma^{1,1} \oplus \Gamma_{E_8}(-1) \oplus \Gamma^{1,1}_g \big)\ , 
\eea
where $\Gamma^{1,1}$ is a two-dimensional lattice with signature $(1,1)$ and inner product 
{\footnotesize $\left( \begin{array}{cc} 0&1\\ 1&0\end{array}\right)$},
and $\Gamma_{E_8}(-1)$ has an inner product given by $-1$ times 
the Cartan matrix of the exceptional group $E_8$.
We denote an integral basis $(\omega_A) = (\omega_S,\omega_i, \omega_a)$  of $H^{2}(Y_E,\bbZ)$,
where $\omega_S$, $\omega_i$ and $\omega_a$ are basis elements of the three terms in 
\eqref{EnriquesCohomology2} respectively. 
We already defined the triple intersections $\cK_{ABC}$ in \eqref{triple_inters}.
Using the relation to the underlying $K3\times \mathbb{T}^2$ one shows
that the only non-vanishing intersections are 
\beq \label{Enriques_int}
   \cK_{S12} = \cK_{S21} = 1\ ,\qquad \cK_{Sa b} =-C_{a b}\ ,
\eeq
where in the appropriate basis the inverse $C^{ab}$ of $C_{ab}$ is the Cartan 
matrix of $E_8$ as already mentioned before.
As in section \ref{orirev} we also introduce a basis $(\tilde \omega^A)=(\tilde \omega^S, \tilde \omega^i, \tilde \omega^a)$ 
of $H^{4}(Y_E,\bbZ)$ dual to $\omega_A$. Finally, we will need to introduce 
a real symplectic basis $(\alpha_A,\beta^A)$ of the third cohomology $H^{3}(Y_E,\bbZ)$.

The explicit form \eqref{EnriquesCohomology2} and \eqref{EnriquesCohomology3} of the integral 
cohomology of $Y_E$
allows us to read of the dimensions $h^{(p,q)}$ of the cohomologies $H^{(p,q)}(Y_E)$.
We find that 
\beq \label{dimHodge}
   h^{(1,1)}(Y_E)= h^{(2,1)}(Y_E)=11\ .
\eeq
This implies that the moduli spaces of complex structure deformations  $\cM_{\rm cs}$
as well as of K\"ahler structure deformations $\cM_{\rm ks}$ are both complex eleven-dimensional.
Moreover, one shows that both of these spaces are the coset \cite{FHSV}
\beq \label{csks}
 \cM_{\rm cs/ks}\ =\  Sl(2,\bbR)/U(1)\ \times\ O(10,2)/\big(O(10)\times O(2)\big)\ ,
\eeq
where $O(q,p,\bbR)$ are orthogonal groups with values in the real numbers. 
The identification $\cM_{\rm cs} \cong \cM_{ks}$ arises due to the fact that
the Enriques Calabi-Yau is self-mirror.
In a careful treatment one also finds that these cosets have to be divided by
the discrete symmetry group
\beq
  O_{E}(\bbZ)\ \equiv\ Sl(2,\bbZ) \times O(10,2,\bbZ)\ ,
\eeq
which is a non-perturbative symmetry of string theory on $Y_E$.
 The presence of this discrete factor is of central importance. All 
 functions on $\cM_{\rm cs/ks}$ have to transform covariantly  under 
 $O_{E}(\bbZ)$ to be well defined.
 Furthermore, note that after dividing by $O_E(\bbZ)$ 
 the identification \eqref{csks} is exact and receives no corrections 
 due to world-sheet instantons \cite{FHSV,mp}. As we will discuss next
 this implies that the Enriques Calabi-Yau is a special example 
 with an exact pre-potential cubic in the moduli around the large volume 
 or large complex structure point.
To make this more precise we discuss the geometry of the moduli 
space $\cM_{\rm ks}$ in more detail. Clearly, due to the fact that $Y_E$
is self-mirror the geometry of $\cM_{\rm cs}$ takes a similar form.

Compactifying Type II string theory on the Enriques Calabi-Yau yields an 
effective four-dimensional theory with $\cN=2$ supersymmetry.
In general, the $\cN=2$ scalar moduli space  consists of a special K\"ahler $\cM_{\rm sk}$
times a quaternionic manifold $\cM_{\rm q}$. For the Enriques Calabi-Yau both spaces 
are cosets. Since we are interested in type IIB compactifications 
we find that the complex structure deformations are the space 
$\cM_{\rm sk}$ while the K\"ahler structure deformations sit inside
the quaternionic space $\cM_{\rm q}$.
One finds \cite{FHSV}
\beq \label{moduli_space}
  \cM_{\rm sk} = \cM_{\rm cs}\ ,\qquad \qquad \cM_{\rm q} = O(12,4)/ \big(O(12)\times O(4)\big)\ \supset\ \cM_{\rm ks} \ .
\eeq
Note that $\cM_{\rm sk}$ is exact and receives no perturbative corrections or corrections due 
to world-sheet or D-instantons. In contrast,  $\cM_{\rm q}$ is in general perturbatively and non-perturbatively 
corrected.
The geometry of the two moduli spaces in \eqref{moduli_space} is encoded by two cubic pre-potentials.
For $\cM_{\rm sk}$ one finds around the large complex structure point 
a pre-potential of the form \footnote{A more careful analysis reveals that there is a linear term $-z^S$ in $\tilde  \cF(z)$ \cite{KM}.
 This term however does not appear in the K\"ahler potential and hence 
not in any physical object discussed in the following.}
\beq \label{def-tildeF}
  \tilde  \cF(z) = - z^S  z^1 z^2 +  \tfrac{1}{2} z^S C_{ab} z^a z^b  \ .
\eeq
Due to the absence of world-sheet instanton corrections this potential is exact and can be transformed 
and used at other points in the moduli space $\cM_{\rm sk}$.
This special K\"ahler manifold encodes deformations of the complex structure through the holomorphic 
$(3,0)$ form
\beq \label{Omega_periods}
  \Omega(z) = X^K(z)\alpha_K - \tilde \cF_K(z)\beta^K\ ,
\eeq
where $(\alpha_K,\beta^K)$ is a real symplectic basis of $H^{3}(Y_E,\bbZ)$. 
The periods of $\Omega$ are thus $(X^K,\tilde \cF_{K})$, 
where $\tilde \cF_{K}$ is the derivative of $\tilde \cF(z)$ with respect to $X^K$. In the 
spacial coordinates $z$ above one has $z^S=X^S/X^0$, $z^i = X^i/X^0$ and $z^a = X^a/X^0$.
One can thus rewrite $\tilde \cF_{K}$ as derivatives with respect to the coordinates $z$ \cite{N=2rev}.

The quaternionic manifold $\cM_{\rm q}$ can be constructed by starting with the underlying special 
K\"ahler manifold $\cM_{\rm ks}(t)$. 
The coordinates $t^A=(S,t^i,t^a)$ are the complexified K\"ahler structure deformations of  $Y_E$ 
arising in the expansion of $-B_2+iJ$ into the two-form basis $\omega_A=(\omega_S,\omega_i,\omega_a)$.
The geometry of the special K\"ahler manifold is determined by the 
pre-potential \footnote{As in \eqref{def-tildeF} we ignore a linear term in $S$ which can be absorbed into a 
redefinition of the coordinates on $\cM_{\rm q}$.}
\beq \label{def-Ft}
    \cF(t)= - S t^1 t^2+\tfrac{1}{2!} S C_{ab} t^a t^b\ .
\eeq
It is straightforward to derive the corresponding K\"ahler potential $ K_{\rm ks}(S,\bar S,t,\bar t) $.
In general, 
$K_{\rm ks}$ can be obtained from the even form $\rho$ introduced in \eqref{def-rho}
by setting $K_{\rm ks} =-\ln i \big< \rho,\bar \rho\big>$ with wedge product defined in footnote \ref{wedge-product}. 
Inserting \eqref{def-Ft} into this expression one evaluates
\beq \label{def-Y}
  K_{\rm ks} = - \ln\big(i(S-\bar S) Y \big)\ , \qquad \quad  Y =   (t-\bar t)^1 (t-\bar t)^2 -\tfrac12 (t-\bar t)^a (t-\bar t)^b C_{a b} \ .
\eeq
The classical quaternionic geometry can be obtained from $\cM_{\rm ks}$ by applying the c-map construction \cite{Ferrara:1989ik}. 
Since our focus will be the orientifold scenario, we will not review the details here. Let us however note
that the quaternionic geometry is invariant under the  K\"ahler transformations of $K_{\rm ks}$. 
It is therefore naturally formulated in terms of the invariant combination $C\rho$, with $C$
proportional to the dilaton $ e^{-\phi}$. Note that $C$ and $\rho$ itself do transform under the K\"ahler 
transformations $K_{\rm ks} \rightarrow K_{\rm ks} -f(t) -\bar f(\bar t)$ as 
\beq \label{trans-Crho}
   C \quad \rightarrow\quad e^{-f} \, C\ ,\qquad \qquad \rho\quad\rightarrow \quad e^f\, \rho\ ,
\eeq 
where $f(t)$ is a holomorphic  function of the moduli.

We will now go one step further and discuss a first set of 
quantum corrections depending on the moduli of 
$\cM_{\rm ks}$. Following \cite{HM,HM2,KM} we will introduce a 
functional $\Phi_{\rm B}$ which counts the leading degeneracies of D(-1), D1, D3
states on the Enriques fiber. Before recalling the precise form of these corrections 
let us note that this investigation will not take place 
in the large volume limit but rather at a second special locus of the 
Enriques moduli space. At this locus also Euclidean D3 branes 
wrapped around a the Enriques fiber are becoming light. To make this more precise, we will choose `dual'
coordinate $\cT^1,\cT^2,\cT^a$ in which large $\I \cT$ implies a small volume 
of the $K3$.
The transformations from the large volume limit 
to this special Enriques locus is given by
\beq \label{def-cT}
   \cT^2 = -\frac{1}{2 t^2}\ ,\qquad \cT^1 = \frac{1}{t^2}\big( t^1 t^2 - \tfrac{1}{2} C_{ab} t^a t^b \big)\ ,\qquad \cT^a = - \frac{1}{t^2} t^a\ .
\eeq
Under this change of coordinates we find that $Y$ defined in \eqref{def-Y} transforms as 
\beq
   2 Y = \frac{1}{2\, \cT^2 \bar \cT^2}\big[2 (\cT -\bar \cT)^1 (\cT -\bar \cT)^2 - (\cT-\bar \cT)^a (\cT-\bar \cT)^b C^D_{a b}  \big]= \frac{1}{\cT^2 \bar \cT^2} Y_D
\eeq
where we have introduced $C^D_{ij} = C_{ij}$, $C^D_{a b} = \tfrac{1}{2} C_{a b}$ and defined $Y_D$. 
In other words, defining the dual K\"ahler potential $K_D(S,\cT)$ as
\beq \label{def-KD}
   K_D(S,\cT) = - \ln \big(i(S-\bar S) Y_D \big) \ ,
\eeq 
one finds that $K_{\rm ks}$ and $K_{D}$
differ only by a K\"ahler transformation.\footnote{%
One finds that $K_{\rm ks}=K_{D} - f -\bar f$,
where $f=- \ln \big(i\sqrt{2}\, \cT^2 \big)$.}
From the coordinate definition \eqref{def-cT} one concludes that the 
corresponding cohomology lattice is 
\beq \label{BHM-reduction}
  \Gamma^{1,1}_s \oplus \Gamma_{E_8}(-2)  \cong H^{0}(E,\bbZ) \oplus H^{4}(E,\bbZ) \oplus \Gamma_{E_8}(-2) 
\eeq
where $H^{0}(E,\bbZ)$ and $ H^{4}(E,\bbZ)$ are the zero and four cohomology of the Enriques 
fiber. This can be seen as follows. The K\"ahler invariant combination to consider is
$C\rho$ with $C$ and $\rho$ transforming as in \eqref{trans-Crho}. One can thus remove the
overall factor of $1/t^2$ in the definitions \eqref{def-cT}. On the one hand this leads to $\cT^2 \propto C$ such 
that $\cT^2$ scales the element in  
$H^{0}(E)$. On the other hand $\cT^1 \propto C(2 t^1 t^2 - C_{ab} t^a t^b)$ which is the 
square of the complexified K\"ahler form and hence parametrizes $H^{4}(E)$. We also 
see that the lattice \eqref{BHM-reduction} contains the self-dual lattice $ \Gamma_{E_8}(-2)$
which has intersection form $C^{D\, a b} = 2 C^{a b}$. The extra factor $2$ arises due 
to the factor $1/2$ in the definition of $\cT^2$. We will see in the next section 
that the coordinates $\cT^1,\cT^2,\cT^a$ have a second advantage, since they can be
identified with the $\cN=1$ coordinates of the orientifold theory. 

We are now in the position to recall a functional $\Phi_{\rm B}(\cT)$ counting 
the leading degeneracies  of Euclidean D(-1), D1, D3 branes on the Enriques fiber.
It was shown in refs.~\cite{borcherdsone, borcherds}, that  for $\cT^i,\cT^a$ with $Y_D < -1$ one 
defines a convergent functional 
\beq \label{Borcherds-Form}
  \Phi_{\rm B} (\cT) = e^{i \cT^1} \prod_{r \in \Pi^+} (1- e^{i r\cdot \cT})^{ (-1)^{m+n} c_{\rm B} (r^2/2)}\ ,
\eeq
where $ r\cdot \cT =  n\cT^1 + m\cT^2 - C^D_{ab} r^a  \cT^b$ for vectors $r=(m,n,r^a)$ in the lattice \eqref{BHM-reduction}.
In the product \eqref{Borcherds-Form} we denote by $\Pi^+$ the set of positive 
roots of the fake monster Lie superalgebra consisting of all nonzero 
vectors $r$  with $r^2 = 2mn-C^D_{ab} r^a r^b \ge -2$ such that $m>0$, or $m=0$ and $n>0$. 
The exponents $c_{\rm B}(r^2/2)$ are given via the modular form 
\beq \label{coeff_funct}
   \sum_n c_{\rm B}(n) q^n = \frac{\eta(q^2)^8}{\eta(q)^8 \eta(q^4)^8}\ ,\qquad \qquad  r^2/2 =n\ ,
\eeq
where $\eta(q)$ is the standard eta function. It was argued in ref.~\cite{KM}
that $ \Phi_{\rm B} (\cT) $ counts the degeneracies of D(-1), D1, D3 branes on the Enriques fiber. To show
this Klemm and Mari\~no \cite{KM} applied a similar argument as Gopakumar and Vafa \cite{GV} 
by performing a Schwinger calculation including the light states at the moduli space locus
parametrized by $\cT^i,\cT^a$. The corresponding BPS particles are bound states of D3 branes wrapping  the Enriques fiber,
D1 wrapped around the curves in the $E_8$ sublattice in \eqref{BHM-reduction} and D(-1) branes.
The leading degeneracies are counted by the lowest genus free-energies $\cF^{(g)}$
of the topological string on $Y_E$. Since $\cF^{(0)}$ is trivial for the Enriques Calabi-Yau 
the first non-trivial contribution arises from a resummation of $\cF^{(1)}$ which precisely 
contains the holomorphic function $ \Phi_{\rm B} (\cT) $. It is important to remark, that $ \Phi_{\rm B} (\cT) $ has particularly 
nice modular properties as we will discuss in section \ref{EnriquesW}. For contributions 
from the higher $\cF^{(g)}$ this is only the case if also a non-holomorphic dependence 
is included. Therefore, we will propose in section \ref{EnriquesW} that $\Phi_{\rm B}$ might contain the leading 
contribution to a holomorphic and modular superpotential of the orientifold theory on the Enriques 
Calabi-Yau.

\subsection{Effective action for the Enriques orientifold \label{effectiveN=1action}}

In this section we study the effective four-dimensional $\cN=1$ supergravity obtained 
by compactifying type IIB supergravity on an orientifold of the Enriques Calabi-Yau $Y_E$.
In order to do this we first have to define an involution $\sigma$ on $Y_E$ and 
investigate its action on the cohomology. 
It was shown in refs.~\cite{Dolgachev, SurfaceB} that involutions on the Enriques surface 
can be characterized by their action on the lattice \eqref{EnriquesCohomology2}.
In particular, there exist an involution acting with a minus sign on 
the $\Gamma_{E_8}(-1)$ term in \eqref{EnriquesCohomology2}, while leaving the $\Gamma^{1,1}$
term invariant. We complete this involution by also inverting the $\bbP^1\cong T^2/\bbZ_2$ base
of the fibration. This keeps the volume form of $\bbP^1$ invariant. We thus find for the 
second cohomology lattice \eqref{EnriquesCohomology2} the split
\beq \label{two_split}
   H^{2}_+(Y_E,\bbZ)\ \cong\ \bbZ \oplus \Gamma^{1,1}\ ,\qquad H^{2}_-(Y_E,\bbZ)\ \cong\ \Gamma_{E_8}(-1)\ ,
\eeq
where $H^2_\pm$ are the plus and minus eigenspaces of $\sigma^*$.
An integral basis $\omega_A=(\omega_S,\omega_i,\omega_a)$ of $H^2(Y_E,\bbZ)$ is introduced 
by setting 
\beq \label{omega_pm}
 \omega_{\alpha}=(\omega_S,\omega_i) \in H^{2}_+(Y_E,\bbZ) \ ,\qquad  \qquad \omega_a \in H^{2}_-(Y_E,\bbZ)\ .
\eeq
This is consistent with the basis $\omega_A$ introduced in the previous section.
The non-vanishing triple intersections $\cK_{Sij}$ and $\cK_{Sab}$
where already given in \eqref{Enriques_int}. It is important to 
note that the orientifold constraints \eqref{inter_constraints} are indeed satisfied, since 
$\cK_{abc}$, $\cK_{a\alpha \beta}$ vanish for $\alpha,\beta$ running over $S,i$.

The odd cohomology $H^3(Y_E,\bbZ)$ also splits into positive and negative eigenspaces 
under the involution. In order to make this split explicit, we note that the above $\sigma$ can be 
extended to the underlying $K3$ surface such that it acts with a minus sign on the $\Gamma_{E_8}(-1)$
terms in the second cohomology lattice $H^{2}(K3,\bbZ)$ given in  \eqref{K3cohom}, while keeping 
the remaining terms invariant. This is of course consistent with the split of the two-cohomology 
\eqref{two_split}. The third cohomology $H^3(Y_E,\bbZ)$ of the Enriques Calabi-Yau 
is obtained by wedging one-forms of the $T^2$ with two-forms of the $K3$ both 
anti-invariant under the $\bbZ_2$ involution defining the Enriques Calabi-Yau. Also including the negative sign 
of $\sigma$ on the two one-forms of $T^2/\bbZ_2$ we thus
find that \eqref{EnriquesCohomology3} splits as
\bea \label{third-cohom}
 H^{3}_+(Y_E,\bbZ)& \cong& \Gamma_{E_8}(-1)\oplus \Gamma_{E_8}(-1)\ ,\\
  H^{3}_-(Y_E,\bbZ) & \cong &  \big(\Gamma^{1,1} \oplus \Gamma_g^{1,1}\big) \oplus \big(\Gamma^{1,1} \oplus \Gamma_g^{1,1}\big) \ . \nn 
\eea
We are now in the position to discuss the reduction of the moduli spaces following 
the general approach in section \ref{orirev}.

Let us first discuss the reduction of the $\cN=2$ special K\"ahler manifold $\cM_{\rm cs}$
spanned by the complex structure deformations $z^\alpha=(z^S,z^i)$ and $z^a$. From \eqref{JOmegatrans} we note 
that the holomorphic three-form $\Omega$ is an element of the negative 
eigenspace of $\sigma^*$. This implies that in the orientifold setup we have $z^a=0$ and 
the expansion \eqref{Omega_periods} reduces to
\bea
   \Omega& =& X^0(\alpha_0 + {z^\alpha}\alpha_\alpha - \tilde \cF_{z^\alpha} \beta^\alpha - (2\tilde \cF - z^\alpha  \tilde \cF_{z^\alpha})\beta^0)\\
         &=&X^0(\alpha_0 + {z^\alpha}\alpha_\alpha + z^1 z^2 \beta^S +  z^S z^2 \beta^1 + z^S z^1 \beta^2 + z^S z^1 z^2\beta^0)\ , \nn
\eea
where $(\alpha_0,\alpha_\alpha,\beta^\alpha,\beta^0)$ is a real symplectic basis of $H^{3}_-(Y_E,\bbZ)$ given in \eqref{third-cohom}.
The prepotential for this reduced special K\"ahler manifold $\tilde \cM_{\rm sk}(z)$ is thus a function of the three moduli 
 $z^\alpha=(z^S, z^i)$ only and takes the form $\tilde \cF(z^I)=- z^S z^1 z^2$.
 The K\"ahler potential is evaluated explicitly to be of the form 
\beq
 K_{\rm cs} =-\ln\big[ i \int \Omega(z) \wedge \bar \Omega(\bar z)\big]
    = -\ln\big[ i (z^S-\bar z^S)(z^1-\bar z^1)(z^2-\bar z^2)\big]\ ,
\eeq
where we have removed the fundamental period $X^{0}$ by a K\"ahler transformation.
The geometry of this reduced moduli space $\tilde \cM_{\rm cs}$ has been studied intensively
in the literature \cite{KM,GKMW}. It can be shown that the mirror map takes a particularly simple 
form due to the absence of world-sheet instantons. It respects the discrete target space 
symmetry $Sl(2,\bbZ)\times \Gamma(2) \times \Gamma(2)$ in the three coordinates $z^S,z^i$ and 
can be given in terms of modular functions of these groups.
Note that in addition to the chiral multiplets just discussed, the projected Enriques theory 
also admits $h^{(2,1)}_+=8$, $\cN=1$ vector multiplets $A_{a}$. The gauge-kinetic coupling 
function has to be holomorphic and is simply given by
\beq
   f_{a b}(z) = -i C_{a b} z^S \ .
\eeq
The kinetic term for $A_{a}$ has coupling matrix $\frac12 \R (f_{a b})=\frac12 C_{a b} \I z^S$ and
is indeed positive definite for $ \I z^S >0$. 

Let us now turn to the discussion of the K\"ahler moduli space $\tilde \cM_{\rm q}$ inside the 
quaternionic moduli space $\cM_{\rm q}$. In \eqref{JBexpansion} and \eqref{CCexpansion} we already 
specified the orientifold invariant expansions of the K\"ahler form $J$, the NS-NS two-form
$B_2$ and the R-R forms $C_2,C_4$. 
In the basis introduced in \eqref{omega_pm} we can summarize these expansions as
\beq
   J = v^S \omega_S + v^i \omega_i \ ,\qquad B_2 = b^a \omega_a\ ,\qquad C_2 = c^a \omega_a\ ,\qquad C_4 = \rho_S \tilde \omega^S+ \rho_i \tilde \omega^i\ ,
\eeq
where the basis $(\tilde \omega^S, \tilde \omega^i)$ of $H^4_+(Y_E,\bbZ)$ is chosen to be dual to $(\omega_S,\omega_i)$.
The real scalar fields $v^a,\rho_a$ as well as $b^S,b^i,c^S,c^i$ have to vanish i.e.~are projected out by the orientifold.
The $\cN=1$ coordinates on the K\"ahler manifold $\tilde \cM_{\rm q}$ are obtained by 
expanding the complex even form $\rho_c$ as in \eqref{def-rhoc}.
This implies that the coordinates $\tau,G^a$ are exactly as given in \eqref{def-tauG}.
The coordinates $T_{\alpha}=(T_S,T_i)$ take the same form as the large volume result 
\eqref{def-T} due to the absence 
of world-sheet instantons in the Enriques Calabi-Yau. Explicitly, one evaluates  
\bea \label{def-TE}
   T_S &=&ie^{-\phi} v^1 v^2 - \tilde \rho_S+\frac{1}{2(\tau -\bar \tau)} C_{a b} G^a (G-\bar G)^b\ ,\\
   T_i &=&  \tfrac{1}{2}i e^{-\phi} v^S v^j- \rho_i\ ,\qquad \qquad i,j=1,2\ , \quad i\neq j\ , \nn 
\eea
where $\tilde \rho_S=\rho_S -\tfrac{1}{2} C_{a b} c^a b^b$. 
The $\cN=1$ K\"ahler potential can be also deduced from our general  
considerations in section \ref{orirev}. More precisely, one uses \eqref{def-TE} together with \eqref{Kqgeneral} or \eqref{Kqgen} to 
evaluate 
\bea \label{def-KqE}
   K_{\rm q} &=& - \ln\big[\tfrac{1}{4}i(T_1-\bar T_1)  \big( 2 (T_S - \bar T_S) (\tau -\bar \tau) - C_{a b} (G-\bar G)^{a} (G-\bar G)^b \big]\nn  \\
   &&- \ln\big[-i(T_2 -\bar T_2)\big]\ .
\eea
This simple explicit form of $K_{\rm q}$ arises 
due to the special form of the intersections \eqref{Enriques_int} 
and the simple cubic pre-potential \eqref{def-Ft}. 
Note that $K_{\rm q}$ is not corrected by $\cN=2$ 
$\alpha'$ contributions, since these vanish identically 
for the Enriques Calabi-Yau.  
In particular, one notices that the perturbative $\alpha'$ corrections proportional to the 
Euler characteristic $\chi(Y_E)$ vanish due to $\chi(Y_E)=2(h^{(1,1)}-h^{(2,1)})=0$. 
We thus conclude that the $\cN=1$ Enriques orientifold theory is particularly 
well under control due to the simplicity of the underlying $\cN=2$ theory.
The $\cN=1$ moduli space $\tilde \cM_{\rm q}$ is also a coset, which is evaluated to be 
of the form 
\beq \label{EcMq}
  \tilde \cM_{\rm q} 
                            \  =\  Sl(2,\bbR)/U(1)\ \times\ \Big(  Sl(2,\bbR)/U(1)\ \times\ O(10,2)/\big(O(10)\times O(2)\big) \Big)\ .
\eeq
Remarkably, we find that the original $\cN=2$ special K\"ahler manifold $\cM_{\rm ks}$ given in \eqref{csks}
arises as the second factor of $\tilde \cM_{\rm q}$. Such a phenomenon was already 
studied from a supergravity point of view in refs.~\cite{D'Auria:2004kx}. In the following we 
will discuss this duality in more detail and make contact to the second parametrization 
of $\cM_{\rm ks}$ introduced in \eqref{def-cT}.

Let us now discuss the appearance of the factor $\cM_{\rm ks}$ in \eqref{EcMq} in more
detail. Recall that we introduced in \eqref{def-cT} a special set of coordinates $S,\cT^i,\cT^a$ 
on $\cM_{\rm ks}$. Imposing the orientifold constraints that in the large volume 
coordinates we have $b^S=b^i=v^\alpha=0$ one shows that the $S, \cT$ coordinates 
truncate as 
\begin{align} \label{dual_limit}
  C\cT^1\ & \rightarrow  \ i  e^{-\phi} \big(v^1 v^2 +\tfrac{1}{2} C_{ab} b^a b^b\big)\ , & C\cT^2 \ &\rightarrow \ \tfrac12 i e^{-\phi}\ ,  \\
  C\cT^a \ & \rightarrow  \ -i e^{-\phi} b^a\ , \qquad &CS \ & \rightarrow\ i v^2 v^S\ .\nn
\end{align}
In this evaluation $C$ was used in the gauge associated to the coordinates $\cT^i,\cT^a$.
It differs  by a factor $2v^2$ from its large volume value $e^{-\phi}$ as imposed 
by its transformation property \eqref{trans-Crho}.  
We can now compare the orientifold truncations \eqref{dual_limit} with the definitions \eqref{def-tauG} and  
\eqref{def-TE} of the $\cN=1$ coordinates. The orientifold limit of the $CS,C\cT^i,\cT^a$ are 
precisely the imaginary parts of $\tau,G^a,T_S, T_1$. 
Viewing the $\cN=1$ coordinates as analytic continuation
we can make the following 
identifications
\beq \label{coord_id}
  \cT^1\  \rightarrow  \ T_S\ , \qquad \quad \cT^2 \ \rightarrow \ \tfrac12 \tau\ ,  \qquad \quad
  \cT^a \  \rightarrow  \ G^a \ , \qquad \quad  S \  \rightarrow\  2 T_1\ .
\eeq
Using this map it is easy to check that also the $\cN=1$ K\"ahler potential \eqref{def-KqE} for the 
scalars $\tau,G^a,T_S, T_1$ can be identified with the K\"ahler potential $K_D$ on $\cM_{\rm ks}$
given in \eqref{def-KD}. This clarifies the fact that the special K\"ahler manifold $\cM_{\rm ks}$
arises as the second factor in the $\cN=1$ moduli space \eqref{EcMq}. 
In the next section we will discuss the holomorphic superpotential
and use the duality map \eqref{coord_id} to propose 
explicit expression for $W$ arising from D3 instantons.

\subsection{The D-instanton superpotential \label{EnriquesW}}

In this section we propose a specific D-instanton superpotential for the 
Enriques orientifold. Since our main focus is the dependence of 
$W_{\text{D-inst}}$ on the moduli $\tau,G^a$ we will concentrate on the 
contribution proportional to $e^{in T_S}$. As seen in \eqref{def-TE} only the complex 
coordinate $T_S$ depends on the fields $G^a$ and hence shifts as discussed 
in section \ref{generalsup}. The imaginary part of $T_S$ contains the volume 
form  of the Enriques fiber modded out by the orientifold involution $\sigma$.
If the corresponding four-cycle $\Sigma$ can be extended to the F-theory picture such that it
contributes to the D-instanton superpotential we expect a correction 
of the form 
\beq \label{Ins_potential}
   W_{\text{D-inst}} = \sum_n \Theta_n (\tau,G^a) e^{in T_S}\ . 
\eeq
In this expression we have also included multi-coverings of $\Sigma$ labeled by $n$. A priory
it is not clear that these will contribute and higher $\Theta_n$ might be zero. 

We will now use our intuition from topological string theory on the 
Enriques Calabi-Yau and conjecture a possible form of $W_{\text{D-inst}}$.
Recall that in section \ref{GeometryN=2} we introduced a specific function $\Phi_{\rm B}(\cT^1,\cT^2,\cT^a)$
encoding the lowest order degeneracies of  D3, D1, D(-1) bound states on the 
Enriques Calabi-Yau. In such states, the D3 instanton wraps the Enriques fiber and 
couples to the complex coordinate $\cT^1$, while 
the D1 branes wrap cycles in the $E_8$ lattice of the second cohomology and
couple to complex coordinate $\cT^a$. The D(-1) couple to the complex 
field $\cT^2$ and appear in generic D3, D1, D(-1) bound states. 
Note that these are also the states which can 
appear in the instanton superpotential \eqref{Ins_potential}.
More precisely, using the map \eqref{coord_id} we identify the coordinates 
$\cT^2,\cT^a$ with the orientifold coordinates $\tau,G^a$.  
The fiber volume appears in $\cT^1$ which is identified with $T_S$.
We now expand the function $\Phi_{\rm B}$ given in \eqref{Borcherds-Form} 
in powers of $e^{inT_S}$ as 
\beq  \label{Phiexp}
      \Phi_{\rm B}(T_S,\tfrac12 \tau,G^a)\ =\ \sum_n \theta_n(\tau,G^a)\, e^{inT_S} \ ,
\eeq
which defines the coefficients $\theta_n(\tau,G^a)$.
Our proposal is that the $G^a$ dependence of the D-instanton superpotential \eqref{Ins_potential}
arises through these functions $\theta_n(\tau,G)$. 
In other words, the superpotential arising due to D3 instantons on the Enriques fiber should take the form 
\beq \label{conjW}
   W_{\text{D-inst}} =  A_0 \sum_n \frac{c_n\ \theta_n(\tau,G^a)}{\eta^{10}(\tau)}\ e^{inT_S} \ ,
\eeq
where $\eta(\tau)$ is the standard eta-function and $c_n$ are appropriate numerical coefficients. 
Unfortunately, without the complete F-theory picture we will not be able to check \eqref{conjW}
directly and details might change in an explicit analysis. However, making contact to the 
discussion in section \ref{generalsup} we will discuss in the remainder of this section 
that  the $\theta_n$ have the correct properties to ensure that $W_{\text{D-inst}}$ is 
a modular form of weight $-1$ in $\tau$. Moreover, also the 
shifts of $T_S$ given in \eqref{tautransgen} and \eqref{Gshift} are appropriately canceled
by shifts of $\theta_n$ as needed for consistency.

Let us finish this section with some remarks on the properties of 
the functions $\theta_n$ in \eqref{conjW}. These can be determined explicitly 
by expanding the expression for $\Phi_{\rm B}$ in the product representation  \eqref{Borcherds-Form} 
or the corresponding sum representation \cite{borcherdsone,borcherds}. It was shown in ref.~\cite{borcherds}
that $\Phi_{\rm B}$ is an automorphic form of weight $4$. Following the arguments of \cite{EZ,binfinite}
one deduces that the coefficient functions $\theta_n$ are Jacobi forms of weight $4$ and index $n$, i.e.~transform
as given in \eqref{transtheta1} and \eqref{transtheta2} under modular transformations and B-shifts. 
In fact, in ref.~\cite{binfinite}  automorphic 
forms similar to $\Phi_{\rm B}$ were constructed by combining 
appropriate Jacobi forms with the exponential $e^{in T_S}$. The precise form 
of $\theta_n$ is then determined by a lift of the modular coefficient functions 
such as \eqref{coeff_funct}.
Instead of giving the explicit expressions for $\theta_n(\tau,G)$ we indirectly 
check some of their properties through a differential equation which they obey.
In order to do that, we note that $\Phi_{\rm B}(\cT)$ satisfies a wave equation of the form \cite{borcherdsone,borcherds}\footnote{This is far from obvious in the product representation of $\Phi_{\rm B}$, but can be easily checked 
when writing $\Phi_{\rm B}$ as a sum \cite{borcherdsone,borcherds}.}
\beq
  2 \frac{\partial^2 \Phi_{\rm B}}{\partial \cT^1 \partial \cT^2}  -
   C_D^{a b}  \frac{\partial^2 \Phi_{\rm B} }{\partial \cT^a \partial \cT^b} =0\ .
\eeq
This equation is readily translated into a condition on the functions $ \theta_n (\tau,G) $ in \eqref{conjW}. 
One finds 
\beq
   \Big(i n \frac{\partial}{\partial \tau} - \tfrac{1}{2} C^{a b}  \frac{\partial^2}{\partial G^a \partial G^b}\Big)   \theta_n (\tau,G)  = 0\ ,
\eeq
which is the higher-dimensional analog of the heat equation for theta-functions
on an appropriate lattice. It also indicates that $\theta_n(\tau,G)$ are Jacobi forms
as expected from the general discussion above. 
Since $\Phi_{\rm B}$ and hence $\theta_n(\tau,G)$ are of weight $4$ we conclude that the inclusion 
of the $\eta^{10}(\tau)$ factor ensures that $W_{\text{D-inst}}$ is of weight 
$-1$ as needed for \eqref{transWundertau}.
To actually show that $\theta_n(\tau,G)$ and $\eta(\tau)$ appear in the correct 
way in the conjectured superpotential \eqref{conjW} one might calculate 
$W_{\text{D-inst}}$ in a specific limit. In particular, it would be interesting to derive $W_{\text{D-inst}}$
in the orbifold limit using its heterotic dual.

\section{Conclusions}

In this paper we discussed the symmetries and non-perturbative 
corrections of the four-dimensional effective theory 
arising in type IIB orientifolds with O3 and O7 planes.
We studied both the K\"ahler potential and 
superpotential in the orientifold large volume limit for general  $\cN=1$ compactifications
and later concentrated on a specific orientifold of the Enriques Calabi-Yau.

In our general analysis we first discussed the $\cN=1$ K\"ahler 
potential including perturbative and non-perturbative $\alpha'$ corrections 
inherited from the underlying $\cN=2$ theory. A subset of the non-perturbative 
$\alpha'$ corrections were shown to survive the 
orientifold large volume limit, since they depend on the 
scalars $G^a$ arising from the NS-NS and R-R two-forms. They
contribute to the K\"ahler potential in an explicitly calculable way, but do not
 alter the $\cN=1$ chiral coordinates.
It was argued that in order to 
ensure duality invariance of the $\alpha'$ corrections to the 
K\"ahler potential also contribution due to D(-1) and D1 branes have to 
be taken into account. In general, it seems hard to determine these 
corrections directly. We thus restrained ourselves to a brief discussion of 
candidate modular completions proposed for the underlying $\cN=2$
theory. It would be interesting to derive these corrections explicitly 
by using heterotic-F-theory duality or be analyzing specific orbifold 
examples. Already the inclusion of the $\alpha'$ corrections will lead 
to interesting new phenomenological properties of these compactifications
and a study of explicit examples is desirable. 

From a phenomenological point of view the two-form scalars $G^a$
have to be rendered massive in a vacuum. 
We have shown that this can be achieved by a potential induced by
D3 instantons. More precisely, we have used the symmetries 
of the orientifold theory to argue that the two-form scalars arise through 
Jacobi forms in front of the D3 instanton contribution $e^{i n T}$ in the superpotential. 
These are generalizations of the well known theta-functions and depend 
on the dilaton-axion $\tau$ as modular parameter. Due to holomorphicity
and modular invariance one might hope that the set of candidate 
Jacobi forms can be restricted to a finite set for a given example. Candidate 
forms should appear in topological string theory on the underlying 
Calabi-Yau manifold counting degeneracies of D1, D(-1) states on cycles which 
become singular in the orientifold background. Additional boundary conditions
obtained in computations performed in specific limits of the theory might then 
fix the precise form of the D-instanton superpotential.

In the finial part of the paper we studied a specific example. We considered an
orientifold of the Enriques Calabi-Yau. The kinetic terms of the 
four-dimensional $\cN=1$ effective theory are determined in terms 
of a simple K\"ahler potential. We showed that the corresponding 
moduli of bulk moduli fields is a product of cosets. Interestingly, the reduction 
of the underlying quaternionic $\cN=2$ geometry led to a K\"ahler manifold which can be identified
with the original deformation space of the complexified K\"ahler structure of the 
underlying Calabi-Yau manifold times an $Sl(2,\bbR)/U(1)$ factor. This duality 
can be used in the study of the D-instanton superpotential on the Enriques Calabi-Yau.
We mapped Jacobi forms known from topological string theory 
on the Enriques Calabi-Yau to the corresponding $\cN=1$ orientifold. This lead to 
a conjecture of a specific D3-instanton superpotential. Unfortunately, explicit 
tests of this proposal are still missing and would involve a careful construction of 
an F-theory realization of the Enriques scenario. It would be also interesting to 
investigate other examples. Particularly, other K3 fibrations might allow to investigate
similar questions, which can then be tested using string-string dualities.

\section*{Acknowledgments}

I would like to thank  Ian Ellwood, Jan Louis, Peter Mayr, Marcos Mari\~no, Liam McAllister, 
Frank Saueressig, Sav Sethi and Gary Shiu for helpful 
comments and discussions. I am particularly grateful to Albrecht Klemm for many 
illuminating discussions and remarks on the draft.
This work was supported in part by NSF CAREER Award No. PHY-0348093, 
DOE grant DE-FG-02-95ER40896, a Research Innovation Award and a Cottrell 
Scholar Award from Research Corporation. 

\bigskip \bigskip

\appendix

\noindent {\bf \Large Appendices}
\section{On the Geometry of the Enriques Calabi-Yau \label{EnriquesGeom}}

In this appendix we review some facts about the geometry of the Enriques 
Calabi-Yau and its cohomology lattice.
Recall the cohomology lattice of the ${\rm K3}$ surface is an even self-dual lattice with Lorentzian
signature. Explicitly, it takes the form \cite{Aspinwall:1996mn}
\bea \label{K3cohom}
   H^2(\bbZ) &\cong& [\Gamma^{1,1}\oplus \Gamma_{E_8}(-1)]_1
\oplus[\Gamma^{1,1}\oplus \Gamma_{ E_8}(-1)]_2\oplus \Gamma^{1,1}_g \ ,\nn \\
   H^0(\bbZ) \oplus H^4(\bbZ)& \cong& \Gamma^{1,1}_s\ ,
\eea
where the inner products on the sublattices $\Gamma_{E_8}(-1)$ and $\Gamma^{1,1}$ are given by   
\beq \label{defC}
  -(C^{a b})=-C_{E_8}\ ,\qquad \qquad (C^{ij})= \left(\begin{array}{cc}0 & 1\\ 1 & 0\end{array} \right)\ . 
\eeq
with $a,b=1,\ldots,8$ and $i,j=1,2$. 
Here $C_{E_8}$ is the Cartan matrix of the exceptional group $E_8$.
In other words, choosing a basis $\tilde \omega_K \in H^{2}(K3,\bbZ)$ with $K=1,\ldots, 22$
one has 
\beq
  \int \tilde \omega_K \wedge \tilde \omega_L = d_{KL}\ ,
\eeq
where $d_{KL}$ equals to $C_{ab}$ on elements of $\Gamma_{E_8}(-1)$ and $C_{ij}$ on 
elements of $\Gamma^{1,1}$ and vanishes for all off-diagonal combinations in the lattice \eqref{K3cohom}.
Clearly, for the torus $\mathbb{T}^2$ we simply have the additional two-dimensional 
lattices $H^1(\mathbb{T}^2,\bbZ)$ and $H^{0}(\mathbb{T}^2,\bbZ) \oplus H^{2}(\mathbb{T}^2,\bbZ)$.
In order to mod out the Enriques involution it is convenient to us an explicit algebraic realization
of the $K3$ surface. For example, a $K3$ surface admitting such an involution can 
be obtained as a double covering of $\mathbb{P}^1 \times \mathbb{P}^1$ branched at the
vanishing locus of a bidegree $(4,4)$ hypersurface \cite{SurfaceB}. The Picard lattice 
of the resulting $K3$ has rank $18$. Using this algebraic realization the action of the 
Enriques involution can be evaluated explicitly.
Let us denote $(p_1,p_2,p_3) \in H^{2}(K3,\bbZ)$ corresponding to the three terms in \eqref{K3cohom}
and abbreviate $p_4 \in H^0(K3,\bbZ) \oplus H^4(K3,\bbZ)$ as well as $p_5 \in H^1(\mathbb{T}^2,\bbZ)$.
The $\bbZ_2$ involution on the Enriques Calabi-Yau 
acts on the elements $p_i$ as~\cite{FHSV} \footnote{The  effect of the phase factor 
on the type II side was interpreted as turning on a Wilson line~\cite{FHSV}.}
\begin{equation}
|p_1,p_2,p_3,p_4,p_5\rangle\rightarrow  e^{\pi i \delta\cdot p_4} |p_2,p_1,-p_3,p_4,-p_5\rangle\ ,
\end{equation} 
where we denoted $\delta=(1,-1)\in \Gamma^{1,1}_s$.                
It it now straight forward to deduce the cohomology of the Enriques 
Calabi-Yau 
\bea \label{EnriquesCohomology2App}
   H^{2}(Y_E,\bbZ) &\cong& \bbZ \oplus \Gamma^{1,1} \oplus \Gamma_{E_8}(-1)\ ,\\
  \label{EnriquesCohomology3App}
   H^{3}(Y_E,\bbZ) &\cong& \big( \Gamma^{1,1} \oplus \Gamma_{E_8}(-1) \oplus \Gamma^{1,1}_g \big)\oplus \big( \Gamma^{1,1} \oplus \Gamma_{E_8}(-1) \oplus \Gamma^{1,1}_g \big)\ , 
\eea
where elements of $H^{2}(Y_E,\bbZ)$ are of the form $p_1 + p_2$ while 
elements of  $H^{3}(Y_E,\bbZ) $ are of the form $p_5 \wedge (p_1 - p_2)$.
One thus shows that the dimensions $h^{(p,q)}$ of the cohomologies $H^{(p,q)}(Y_E)$ are
$h^{(1,1)}(Y_E)= h^{(2,1)}(Y_E)=11$.
The Enriques Calabi-Yau is shown to be self mirror \cite{FHSV}. The two eleven-dimensional 
moduli spaces of complex and K\"ahler structure deformations are identified with the coset \eqref{csks}
mod the symmetry group $Sl(2,\bbZ)\times O(10,2,\bbZ)$ as discussed.


\end{document}